\documentclass{JHEP3}
 \usepackage{epsfig}

 \title{
  Local charge compensation from colour preconfinement as a key to the
dynamics of hadronization}

 \author{Kosuke Odagiri\\
  Institute of Physics, Academia Sinica, Nankang, Taipei, Taiwan 11529,
  The Republic of China}

 \abstract{
  If, as is commonly accepted, the colour-singlet, `preconfined',
perturbative clusters are the primary units of hadronization, then the
electric charge is necessarily compensated locally at the scale of the
typical cluster mass. As a result, the minijet electric charge is
suppressed at scales that are greater than the cluster mass.
  We hence argue, and demonstrate by means of Monte Carlo simulations
using HERWIG, that the scale at which charge compensation is violated is
close to the mass of the clusters involved in hadronization, and its
measurement would provide a clue to resolving the nature of the dynamics.
  We repeat the calculation using PYTHIA and find that the numbers 
produced by the two generators are similar.
  The cluster mass distribution is sensitive to soft emission that is
considered unresolved in the parton shower phase. We discuss how the
description of the splitting of large clusters in terms of unresolved
emission modifies the algorithm of HERWIG, and relate the findings to the
yet unknown underlying nonperturbative mechanism.
  In particular, we propose a form of $\alpha_S$ that follows from a
power-enhanced beta function, and discuss how this $\alpha_S$ that governs
unresolved emission may be related to power corrections.
  Our findings are in agreement with experimental data.
  }


 \preprint{hep-ph/0307026}

\begin{document}

 \section{Introduction}

  Hadronization is one of the most poorly understood aspects of QCD.
  Our knowledge is currently limited to models \cite{herwig,pythia} and
estimations of the power-suppressed corrections \cite{power} to
observables that have sensitivity to soft physics.

  Of the little that is known about the dynamics of hadronization, colour
preconfinement \cite{amati_veneziano,preconfinement}, which is a general
property of perturbative QCD, is often regarded as a plausible starting
point.

  Colour preconfinement is a theorem, which follows from perturbative QCD,
that states that in the course of the evolution between the hard scale
$Q_{\mathrm{hard}}$ and the cut-off scale $Q_0$ that results in the
formation of a perturbative parton shower, the quarks and gluons become
organized in colour-singlet `clusters', whose mass is of order $Q_0$ and
is independent of $Q_{\mathrm{hard}}$ in the limit of large
$Q_{\mathrm{hard}}$.

  It has been proposed \cite{amati_veneziano} that the clusters so
produced participate independently in hadronization. If so, and provided
that the physical cut-off scale to the parton shower is found to be small
compared with $Q_{\mathrm{hard}}$, the nonperturbative contribution to
jets should not dramatically disturb the properties of the perturbative
parton shower.

  Models of hadronization based on colour preconfinement, notably the
Monte Carlo event generators HERWIG \cite{herwig} and PYTHIA
\cite{pythia}, have been found to agree, to within creditable accuracy,
with experimental data. When carrying out this comparison, since the
dynamics of hadronization is little understood, it is necessary to tune
some parameters that are related to hadronization.

  One of the key parameters is the cut-off scale $Q_0$, above which it is
appropriate to apply the perturbative parton shower evolution and below
which nonperturbative dynamics dominates.

  By means of experimental tunes it has been discovered that the cut-off
scale is quite low, $\mathcal{O}$(1 GeV). Thus the typical cluster mass
is also small.

  Despite the success of the models of hadronization based on
preconfinement, the existence of these preconfined clusters hitherto lacks
direct experimental evidence\footnote{However, we note that the minijet
structure of jets has been studied in the context of cluster hadronization
in ref.~\cite{cdf_subcluster}}.
  In view of this, we turn our attention to the phenomenon of local charge
compensation \cite{local_charge_compensation}. We demonstrate that colour
preconfinement naturally leads to local charge compensation.
  We advocate the measurement of minijet charge rather than distribution
in rapidity space which has been traditionally considered.
  This change of observable allows us to relate the scale of charge
compensation to the scale of colour preconfinement.

  In HERWIG, the perturbative clusters that remain large at the end of the
parton shower phase are split according to a parametrization. We point out
that because the mass distribution of the clusters is sensitive to soft
emission that is considered unresolved in the parton shower phase, this
cluster-splitting dynamics may be rephrased in terms of unresolved
emission governed by a modified low energy running strong coupling. By
comparing the result with the default cluster mass distribution of HERWIG,
we can estimate the energy scale involved in the cluster-splitting phase,
or the shape of the modified $\alpha_S$, and derive a possible
phenomenological distinction.

  The cluster splitting energy scale thus established may be interpreted
loosely as the scale of `emission before confinement'. We discuss one
explicit interpretation where enhanced $g\to q\bar q$ splitting modifies
the running of $\alpha_S$. The resulting form of low-energy $\alpha_S$
agrees with our findings with physically acceptable parameter values.
 We then discuss how this $\alpha_S$ relates to the part of the power
corrections that is due to soft gluon emission. As the $\alpha_S$ has
complex poles, there is ambiguity associated with its analytization.

  This paper is organized as follows. We first introduce the concept of
colour preconfinement and the resulting local compensation of charge. We
proceed to define the relevant observables. We carry out simulations using
the HERWIG Monte Carlo event generator and compare the numbers against
those of PYTHIA. We discuss the cluster-splitting procedure in HERWIG that
affects this observable, and consider possible physics interpretations and
consequences. Conclusions are stated at the end.

 \section{Colour preconfinement and charge compensation}

  In the original proposal of Amati and Veneziano \cite{amati_veneziano},
units composed of perturbatively emitted quarks and gluons become
colour-singlet clusters.  Obviously the `ends' of these clusters are
defined by quarks. However, as the emission of quarks in a parton shower
is a relatively rare event, the resultant clusters can be quite large even
though the mass is still determined by $Q_0$ in the limit of large
$Q_{\mathrm{hard}}$.
  In the Lund string model of PYTHIA, these quark-gluon systems are
regarded as `kinked strings' that subsequently decay into hadrons.

  An alternative, and more economical, approach adopted in the cluster
hadronization model \cite{cluster_hadronization} of HERWIG is to introduce
forced, `nonperturbative', splitting of gluons at the end of the parton
shower, such that all clusters are quark-antiquark colour-singlet dipole
systems. However, even in this case, there often remain in the end some
clusters that are considered too large to be nonperturbative objects.
These clusters are then split by a power distribution in terms of the
masses of the decay products.
  We shall discuss this in more detail in
secs.~\ref{sec_herwigsimulations} and \ref{sec_interpretation}.
  The small clusters at the end decay isotropically into hadrons 
according to the phase space weight.

  We may therefore say that the primary units of hadronization are, in the
former picture, perturbative colour-singlet systems, whereas in the
latter, all colour connected two-parton systems, or in other words,
dipoles.

  In any case, if the physical cut-off scale is at
$Q_{0\mathrm{(physical)}}$, then for all artifical cut-off scales $Q_0$
greater than $Q_{0\mathrm{(physical)}}$, emitted partons are organized
into clusters with mass of order $Q_0$. Therefore if we take an observable
that is insensitive to physics below $Q_0$, we would be measuring the
properties of clusters with mass of order $Q_0$.

  The electric charge of a cluster is always 0 or $\pm1$ because a cluster
is in effect a quark-antiquark system. Following our reasoning above, the
charge is 0 or $\pm1$ at all scales above $Q_{0\mathrm{(physical)}}$, such
that if a cluster at a certain scale is composed of a number of smaller
clusters at a smaller scale, the electric charges of these smaller
clusters must mutually cancel.

  Hence electric charge is `locally' compensated if colour is preconfined,
rather than increasing in proportion to the square root of the number of
charged tracks belonging to the cluster as would be expected if the
charges were uncorrelated.
  It is worth noting that the converse is not necessarily true. In
particular, in PYTHIA, although the strings may be quite large, their
decay proceeds by the creation of quark-antiquark pairs, such that the
scale at which charge is compensated is in general small compared to the
mass of the strings.

  At scales lower than $Q_{0\mathrm{(physical)}}$, on the other hand, if
nonperturbative effects are dominant, there is no reason to expect that
charge is locally compensated. This is illustrated in 
fig.~\ref{fig_compense}.

 \EPSFIGURE[ht]{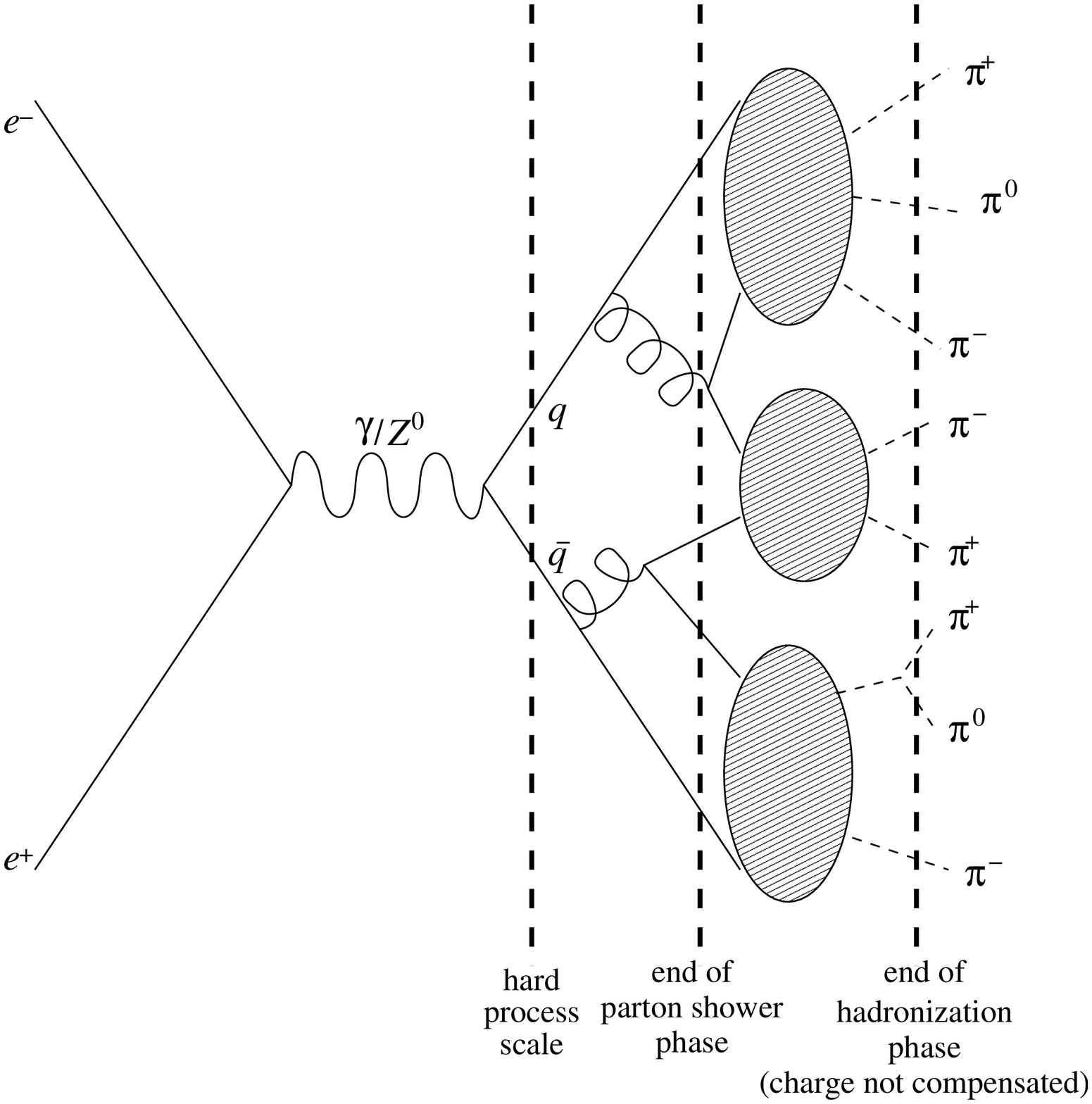,width=14cm}
 {An illustration of the violation of local charge compensation during
hadronization in the colour preconfined picture (HERWIG). Neighbouring
charges do not necessarily cancel.\label{fig_compense}}

 \section{Charge compensation observables}\label{sec_observables}

  In practice, it is not possible to define an exclusive observable that
is absolutely independent of physics below a certain arbitrary scale. 
However, we would physically expect that a minijet charge, if 
appropriately defined, can minimize the contamination.

  The sensitivity to physics below the scale defined by the minijet
resolution variable depends on the algorithm that is used to combine
tracks or objects consisting of tracks. We expect that $k_T$ based
algorithms \cite{durham_algorithm,cambridge_algorithm} exhibit the most
desirable properties since the resolution variable $k_T$ defined between
two objects is, in principle, the energy scale that is required for a
splitting that creates the two objects. In view of this, let us as our
default procedure adopt the Cambridge \cite{cambridge_algorithm}
algorithm. This algorithm adopts as the test variable between objects
$(ij)$ a variable that is essentially $k_T^2$:
 \begin{equation}
  y_{ij}\propto \min(E_i,E_j)^2v_{ij}.
  \label{eqn_yij}
 \end{equation}
  The normalization factor is taken to be $1/E^2_\mathrm{vis}$ where
$E_\mathrm{vis}$ is the total visible energy.
  The quantity $v_{ij}$, which also serves as the ordering variable, is
defined by:
 \begin{equation}
  v_{ij}=2(1-\cos\theta_{ij}).
 \end{equation}
  The jet construction algorithm is iterative and the pair $(ij)$ with the
smallest value of $v_{ij}$ that satisfies $y_{ij}<y_\mathrm{cut}$ is
combined. In addition, for a pair which does not satisfy
$y_{ij}<y_\mathrm{cut}$ but have smaller values of $v_{ij}$, the object
with lower energy is `frozen' by regarding this object as a jet. This last
point marks the distinction between the Cambridge algorithm and the
`angular-ordered Durham' algorithm \cite{cambridge_algorithm}.

  We now define the minijet charge as
 \begin{equation}
  \mathrm{Q}_\mathrm{minijet}(y_\mathrm{cut})=
  \sum \mathrm{Q}_\mathrm{track}.
 \end{equation}
  The summation is over the tracks that belong to the minijet under
consideration.
  It is also instructive to define a charge that is weighted by powers of
the track three-momenta, as:
 \begin{equation}
  \mathrm{Q}_\mathrm{minijet}(y_\mathrm{cut},\kappa)=
  \frac{
  \sum
  \mathrm{Q}_\mathrm{track}
  \left(\mathrm{p}_\mathrm{track}\cdot\mathrm{p}_\mathrm{minijet}\right)
  ^\kappa
  }{
  \sum
  \left(\mathrm{p}_\mathrm{track}\cdot\mathrm{p}_\mathrm{minijet}\right)
  ^\kappa
  }, \label{eqn_kappajet}
 \end{equation}
  so that comparison could be made with literature \cite{aleph_ab}.
  The $\kappa\to0$ limit of
$\mathrm{Q}_\mathrm{minijet}(y_\mathrm{cut},\kappa)$ differs from the
unweighted charge $\mathrm{Q}_\mathrm{minijet}(y_\mathrm{cut})$ by a
factor of track multiplicity.

  Using the charge thus defined, we may further define the following
quantities. First, the average minijet charge in an event is defined as:
 \begin{equation}
  <|\mathrm{Q}_\mathrm{minijet}(y_\mathrm{cut},\kappa)|>=
  \frac{
  \sum
  |\mathrm{Q}_\mathrm{minijet}(y_\mathrm{cut},\kappa)|
  }{\#_\mathrm{minijets}}. \label{eqn_kappajets}
 \end{equation}
  The measurement of the average minijet charge, measured and averaged
over a sample of events, can confirm that charge is locally compensated.
On the other hand,
  it would also be interesting to investigate to what extent the charge is
limited to zero or $\pm1$. We hence define the ratio:
 \begin{equation}
  R_{\mathrm{|Q|}>1}=\frac{
  \#_{\mathrm{|Q|}>1}
  }{\#_\mathrm{minijets}}.
 \end{equation}
  An alternative possibility would be to consider the average of some
powers of minijet charge.

 \section{Result of simulations}\label{sec_herwigsimulations}

  We present the result of HERWIG simulations, and later provide a
comparison with results obtained using PYTHIA, for the observables defined
above. Unless stated otherwise, the simulations are at the hadron level
and the default values in HERWIG 6.500 and PYTHIA 6.215 are adopted for
the parameters affecting hadronization. The HERWIG parameters are:
 \begin{eqnarray}
  \mathtt{RMASS(13)}=0.75, \qquad \mathtt{CLMAX}=3.35, &&\qquad
  \mathtt{CLPOW}=2.0,\label{eqn_clmax}\\
  \mathtt{RMASS(1)}=\mathtt{RMASS(2)}&&=0.32,\\
  \mathtt{PSPLT(1)}=\mathtt{PSPLT(2)}&&=1.0,\\
  \mathtt{CLDIR(1)}=\mathtt{CLDIR(2)}&&=1,\\
  \mathtt{CLSMR(1)}=\mathtt{CLSMR(2)}&&=0.
 \end{eqnarray}
  Out of the parameters listed above, the most important are the
following. The effective gluon mass {\tt RMASS(13)} (in GeV) sets the
cut-off scale for the parton shower, {\tt CLMAX} (in GeV) sets the maximum
allowed cluster mass, and {\tt PSPLT} is the power for the mass
distribution in the splitting of the clusters that have masses exceeding
{\tt CLMAX}. The cluster mass distribution before and after the cluster
splitting is shown in fig.~\ref{fig_clusmas}.
  For comparison, we also show the distribution corresponding to {\tt
RMASS(13)}=1.5 GeV, {\tt CLMAX}=5 GeV, which results in larger cluster
masses than the default.

 \EPSFIGURE[ht]{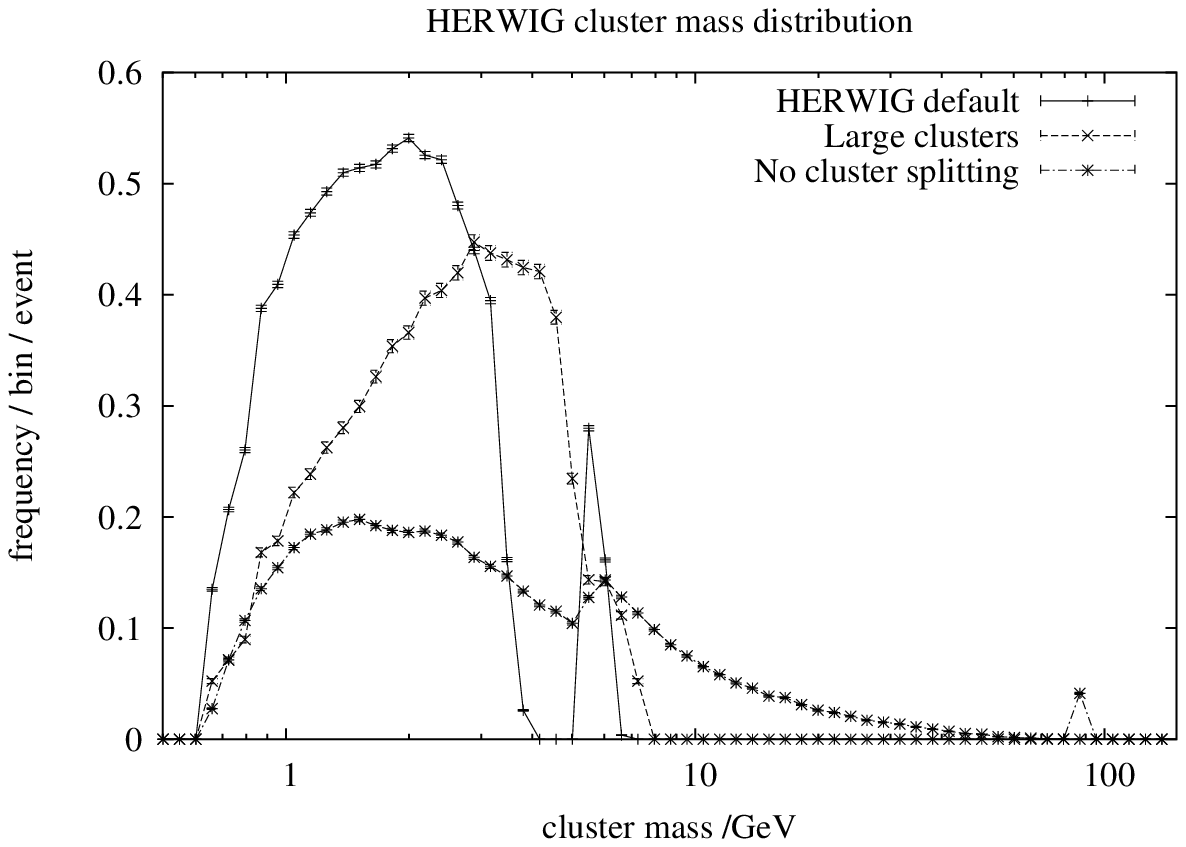,width=14cm}
 {The HERWIG cluster mass distribution. The curve `HERWIG default' is
generated by the default parameter settings.
  The curve `large clusters' corresponds to the choice {\tt
RMASS(13)}=1.5 GeV, {\tt CLMAX}=5 GeV.
 We also show the result of having cluster splitting forbidden by setting
an arbitrarily large {\tt CLMAX}.
  The peak at about 5 GeV is due to clusters containing the bottom quark.
 \label{fig_clusmas}}

  To clarify, when the parton shower is terminated by the effective gluon
mass, clusters are first formed by splitting the gluons into $u\bar u$ or
$d\bar d$. When they are heavier than approximately {\tt CLMAX}, clusters
are split by generating an additional $u\bar u$ or $d\bar d$ pair. This is
performed by uniformly generating mass raised to the power of {\tt PSPLT}.
  At the end, clusters are decayed according to the phase space weight 
into the available hadrons.

  The other parameters, which are less important for our purpose, are the
down- and up-quark masses {\tt RMASS(1)}, {\tt RMASS(2)}, the maximum
cluster mass parameter {\tt CLPOW} and the parameters {\tt CLDIR} and {\tt
CLSMR} that define the extent to which the cluster decay remembers the
direction of perturbatively produced quarks.

  For the main part of this section, we consider the simulation of the
$e^-e^+\to q\bar q$ hard subprocess at the $Z^0$ pole,
$\sqrt{s}=M_{Z^0}=91.188$ GeV, such that
${k_T}_\mathrm{cut}=M_{Z^0}\sqrt{y_\mathrm{cut}}$. Raising the energy
raises the overall multiplicity and therefore raises contamination due to
the mis-identification of charged tracks when constructing minijets.
  The initial/final state photon radiation is switched off in both HERWIG
and PYTHIA. There is no detector simulation.
  The sample size is 10000 events except where otherwise stated.

 \EPSFIGURE[ht]{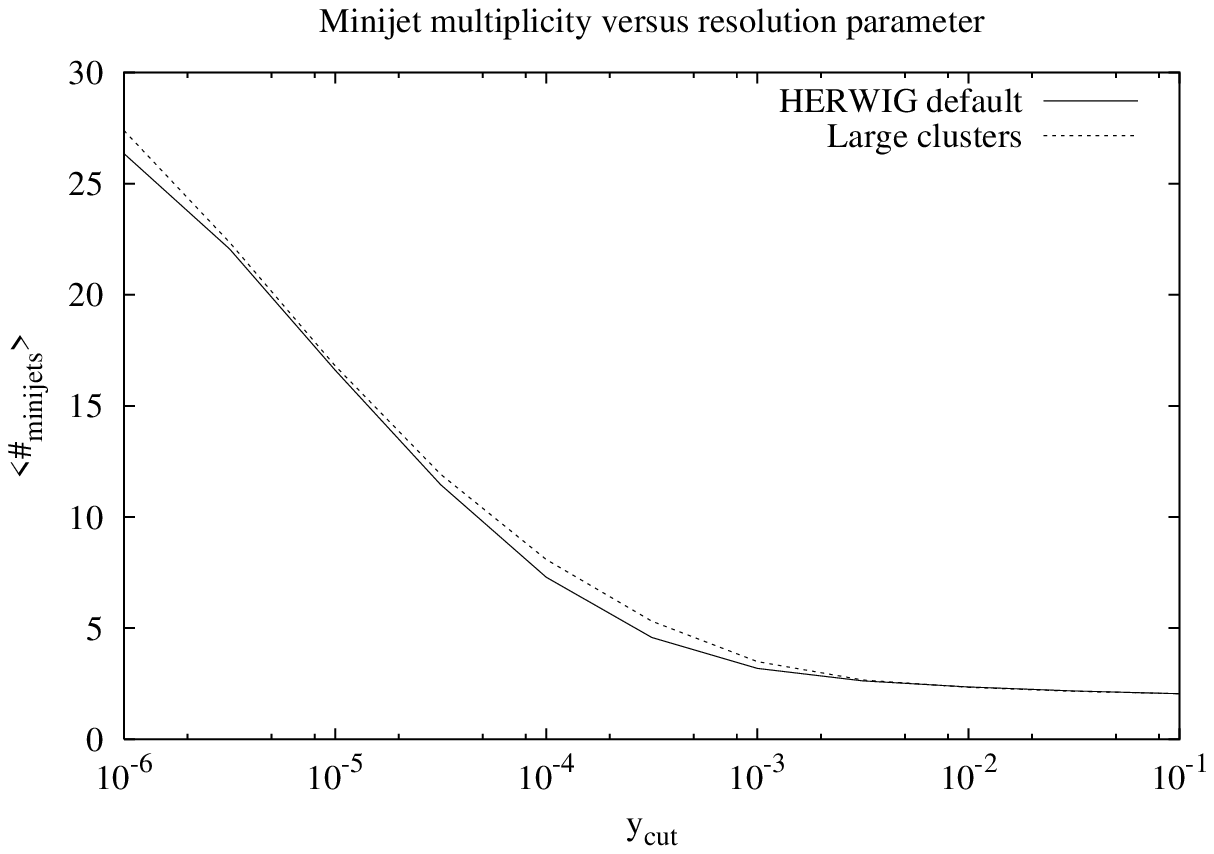,width=14cm}
 {The minijet multiplicity plotted against the Cambridge algorithm
resolution parameter $y_\mathrm{cut}$.
  Sample size is 1000 events. 
  The error due to finite statistics is small.
 \label{fig_multiplicity}}

  We first show the minijet multiplicity in fig.~\ref{fig_multiplicity}.
Towards large $y_\mathrm{cut}$ we see that it tends to 2, whereas for
small $y_\mathrm{cut}$ it begins to saturate as the number of tracks is
finite. A $y_\mathrm{cut}$ value of $10^{-4}$ corresponds to a
${k_T}_\mathrm{cut}$ of 0.91 GeV and hence of order of the cluster mass
scale when using the default parameter values in HERWIG.

  In the same figure, we also show the result of adopting large clusters
({\tt RMASS(13)}=1.5 GeV, {\tt CLMAX}=5 GeV).
  The difference between the two curves is slight, and the physics behind
the difference is unclear. The minijet multiplicity itself is therefore
not a sufficiently good observable for studying hadronization.

  We now turn our attention to the average minijet charge. We carry out
our simulation again for the HERWIG default and modified parameter values.

 \EPSFIGURE[ht]{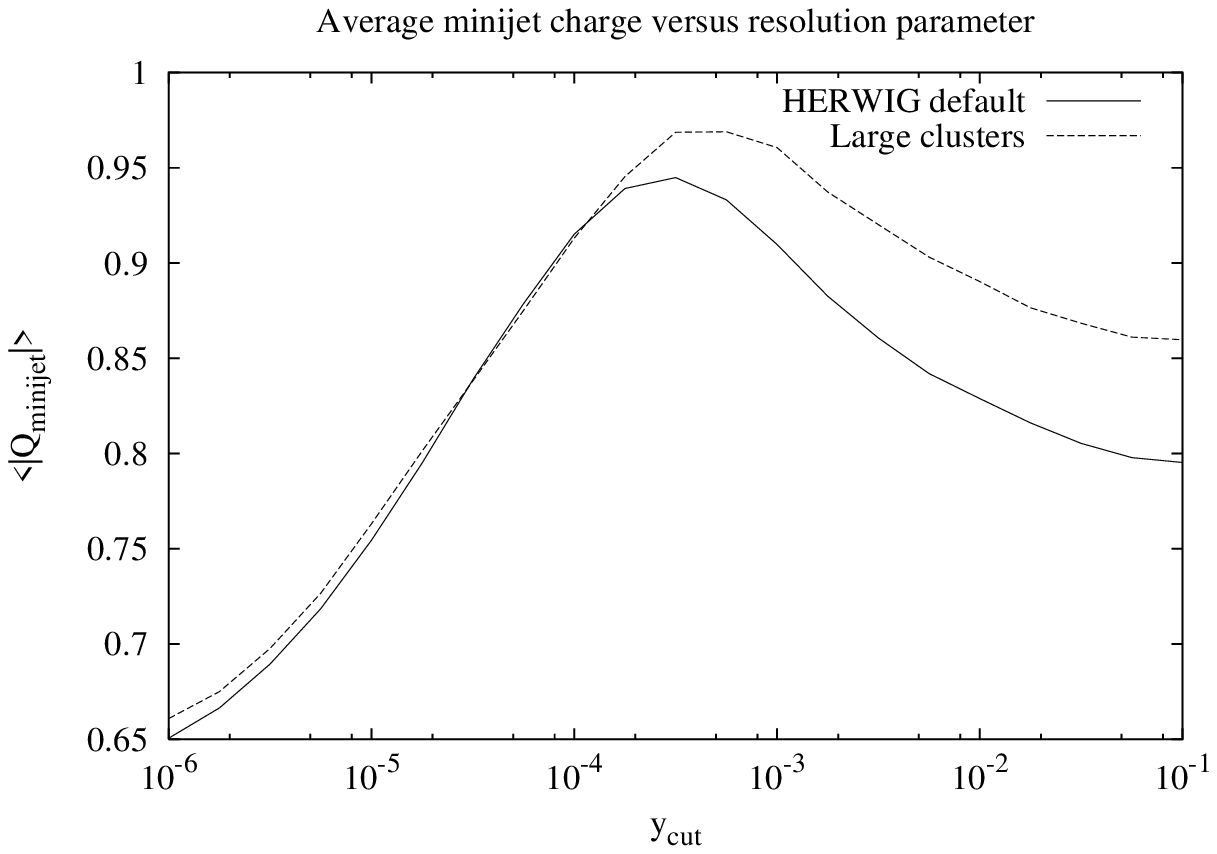,width=14cm}
 {The mean minijet charge plotted against the Cambridge algorithm
resolution parameter $y_\mathrm{cut}$.
 \label{fig_unweightedcharge}}

  The result is shown in fig.~\ref{fig_unweightedcharge}. The Monte-Carlo
sample size is 10000 events, but even with $\mathcal{O}(1000)$ events, the
statistical error is small enough to allow comparison of the two curves.
We now find that in contrast to the minijet multiplicity, we have an
observable whose behaviour can be directly related to the physics of
hadronization. In particular, the peak position of the average minijet
charge observable, where local charge compensation is maximally violated,
is close to the cluster mass scale.
  Far below this scale, charge becomes that of individual hadrons, whereas
above this scale, local charge compensation ensures that the minijet
charge does not increase arbitrarily.

  Another point that is worth noting is that the minijet charge at large
$y_\mathrm{cut}$, even in the two-jet limit, is fairly sensitive to the
cluster mass. If the scale of hadronization is large, minijet charge is
contaminated by the reshuffling of charged tracks amongst minijets. An
alternative way of seeing this is through plotting the fraction of
minijets that have charge exceeding 1, as shown in fig.~\ref{fig_rqgt1}.

 \EPSFIGURE[ht]{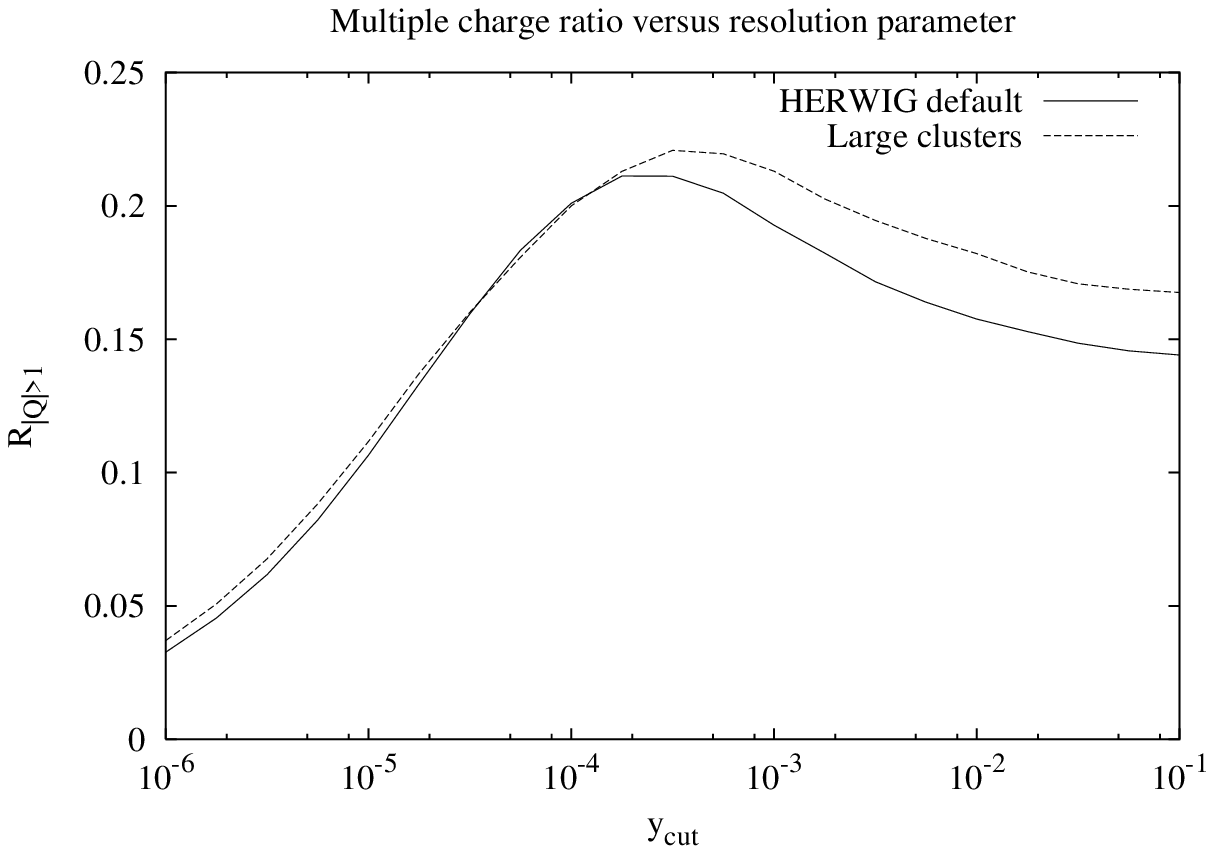,width=14cm}
 {The fraction of minijets that have charge exceeding 1, plotted against
the Cambridge algorithm resolution parameter $y_\mathrm{cut}$.
 \label{fig_rqgt1}}

  The relation between the minijet charge and the multiple-charge fraction
can be understood as follows. Let us first consider the charge of a quark
or a gluon jet in the ideal case where all clusters emitted from the
originating parton is assigned to this jet. In this case, the net charge
of this jet has to be equal to $0$ or $\pm1$ with equal probability. This
is because the charge is determined by the typically nonperturbative
splitting $g\to q\bar q$ that is closest to the hard process. All other
emissions conserve the jet charge. This produced quark has equal
probability of being a $u$ or a $d$. For the case of a quark jet, for
instance, regardless of whether the originating parton is up-type or
down-type, the net charge of the jet is therefore always $0$ or $\pm1$
with equal probability. Hence the `perturbative' jet charge should average
to $0.5$ so long as the charges of different jets are uncorrelated, or
perfectly correlated in the case of the two-jet limit.

  Now let us consider the contamination of this jet with at most one
singly charged track, whose charge is uncorrelated with the charge of the
jet, with equal probability $p$ for each charge, $+1$ or $-1$. We can show
that the probabilities for the total jet charge are now modified to:
 \begin{eqnarray}
  \label{pqeq0}
  P_{\mathrm{|Q|}=0} &=& \frac{1-2p}2+\frac{p}2=\frac{1-p}2, \\
  \label{pqeq1}
  P_{\mathrm{|Q|}=1} &=& \frac{2p}2+\frac{1-2p}2=\frac12, \\
  \label{pqeq2}
  P_{\mathrm{|Q|}=2} &=& \frac{p}2.
 \end{eqnarray}
  Summing the charge times probability for each case, we obtain the
expectation value for the net charge as $p+1/2$. On the other hand, the 
multiple-charge fraction is given by $P_{\mathrm{|Q|}=2}$ above, such that 
we obtain the relation between the minijet charge and the multiple-charge 
fraction as:
 \begin{equation}
  <|\mathrm{Q}_\mathrm{minijet}|>=\frac12+2R_{\mathrm{|Q|}>1}.
  \label{eqn_charge_vs_rqgt1}
 \end{equation}
  Comparing figs.~\ref{fig_unweightedcharge} and \ref{fig_rqgt1}, we see
that this relation is satisfied very well in the two-jet limit.
  For lower values of $y_\mathrm{cut}$, as the nonperturbative effects
become more dominant, the relation is increasingly violated, although the
general behaviour is still in accord with the expectation from
eqn.~(\ref{eqn_charge_vs_rqgt1}).

  The assumption here of contamination due to at most one charged track is
sufficient for our discussion, but for the sake of completeness, let us
derive in outline the case without restriction on the number of
uncorrelated contaminating tracks. We first define the generating function
for the jet charge by:
 \begin{equation}
  \left(\frac1{4u}+\frac12+\frac u4\right)
  \exp\left[p\left(\frac 1u-2+u\right)\right]
  = \sum_{n=-\infty}^{\infty}u^nP_{\mathrm{Q}=n}. \label{generating}
 \end{equation}
  We may obtain this exponential form as the limit of binomial
distribution due to infinitely many contaminating tracks, which is a
direct generalization of eqns.~(\ref{pqeq0})--(\ref{pqeq2}).
  We recover eqns.~(\ref{pqeq0})--(\ref{pqeq2}) in the limit of small $p$.

  Bessel functions of the first kind are defined by the generating
function:
 \begin{equation}
  \exp\left[\frac z2\left(t-\frac 1t\right)\right]
  = \sum_{n=-\infty}^{\infty}t^nJ_n(z).
 \end{equation}
  After a few elementary algebraic manipulations, we can equate the two
generating functions to obtain:
 \begin{equation}
  P_{\mathrm{Q}=n}=e^{-2p}\left[
  \frac14{i^{n-1}J_{n-1}\left(-2ip\right)}+
  \frac12{i^{n  }J_{n  }\left(-2ip\right)}+
  \frac14{i^{n+1}J_{n+1}\left(-2ip\right)}\right].
 \end{equation}
  Bessel functions can be expanded in a series:
 \begin{equation}
  i^mJ_m\left(-2ip\right)=\sum_{l=0}^{\infty}
  \frac{p^{2l+|m|}}{l!(|m|+l)!},
 \end{equation}
  such that corrections to eqns.~(\ref{pqeq0})--(\ref{pqeq2}) can be
obtained systematically as an expansion in $p$. Furthermore, we observe
that by successively operating on eqn.~(\ref{generating}) by $u(d/du)$, we
may obtain the expectation value for charge raised to any even integer
power. For instance, operating twice with $u(d/du)$ we obtain:
 \begin{equation}
  <\mathrm{Q}^2_\mathrm{minijet}>=\frac12+2p.
 \end{equation}

  We can estimate the contamination probability $p$ naively as the ratio
of the cluster mass against the jet energy multiplied by the cluster
multiplicity:
 \begin{equation}
  p\sim \frac{M_\mathrm{clus}}{\sqrt{s}/2}\times\#_\mathrm{clus}.
 \end{equation}
  The cluster multiplicity can be estimated from fig.~\ref{fig_clusmas}.
As the cluster size increases, the cluster multiplicity decreases and so
there is some cancellation between the two contributions. This naive
estimate can be compared with the two-jet limit of fig.~\ref{fig_rqgt1}
and we see that both the cluster mass dependence and the overall magnitude
is described well. Thus the net jet charge in the two-jet limit by itself
already provides a measure of the cluster mass scale.

 \EPSFIGURE[ht]{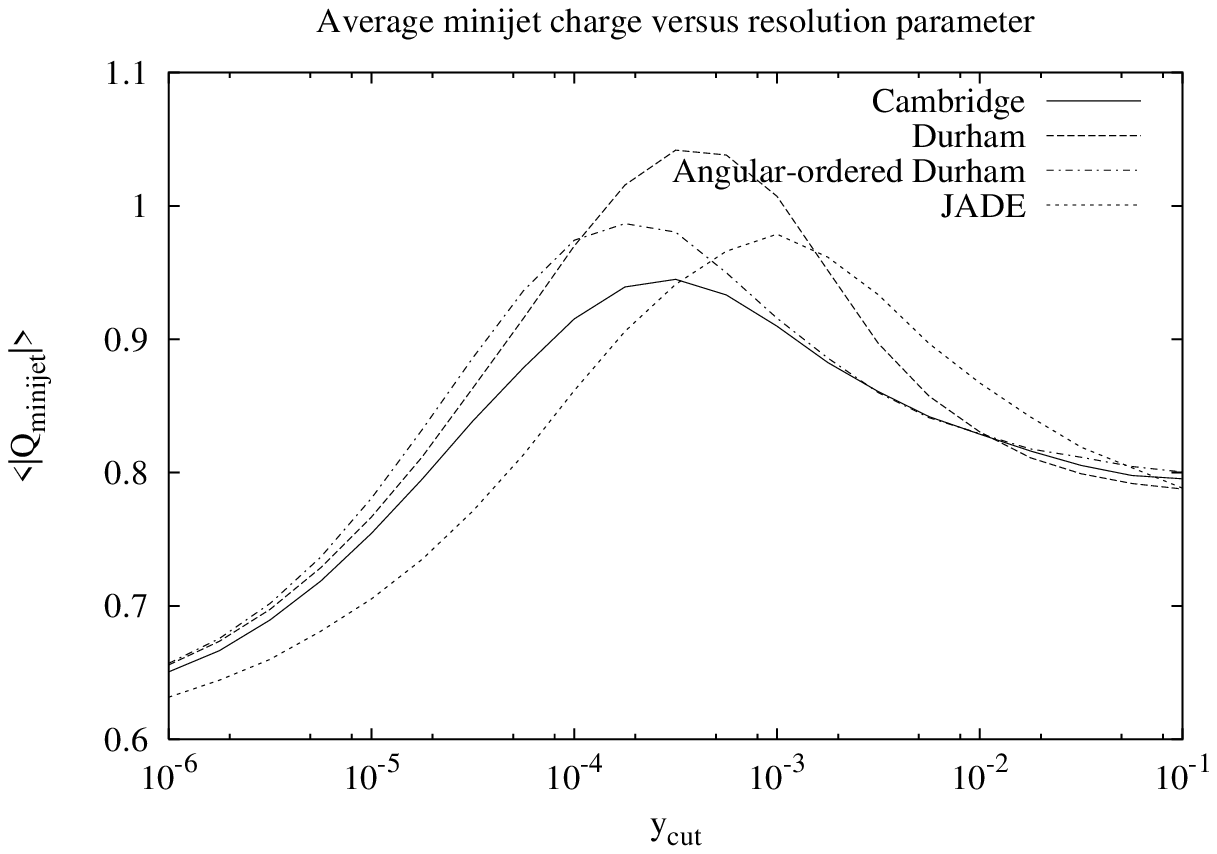,width=14cm}
 {A comparison of the minijet charge obtained using different jet 
algorithms.
 \label{fig_durham}}

  Having said this, it is important to test to what extent the
contamination is a feature of nonperturbative dynamics and not a defect of
the jet algorithm, i.e., misclustering. To this effect, we repeat the
analysis using different jet algorithms and show the result in
fig.~\ref{fig_durham}.

  The JADE algorithm \cite{jade_algorithm}, that uses the invariant mass
rather than $k_T$ as the resolution variable, has peak at higher
$y_\mathrm{cut}$ than the $k_T$-based algorithms. One reason is the
different kinematics, namely that the $k_T$ between two tracks is always
smaller than their invariant mass. Another reason is the increased
mis-clustering at high $y_\mathrm{cut}$ where the dynamics is
perturbative, and reduced mis-clustering at low $y_\mathrm{cut}$ where the
dynamics is nonperturbative.

  Before discussing this point in more detail, we turn our attention to
the three remaining, $k_T$-based, algorithms. The mis-clustering at
moderate $y_\mathrm{cut}$ is largest for the Durham algorithm and is
slightly better for the angular-ordered Durham algorithm that was
mentioned in sec.~\ref{sec_observables}. At very low $y_\mathrm{cut}$,
where angular-ordering is no longer a feature of the relevant dynamics,
the Durham algorithm actually fares better than the angular-ordered
algorithm.

  At large $y_\mathrm{cut}$ the minijet charge due to the four algorithms
converge. In particular, there is little distinction between the large
$y_\mathrm{cut}$ behaviour of the Cambridge algorithm and the
angular-ordered Durham algorithm from which the former algorithm is
derived. Hence it seems reasonable to suppose that the minijet charge
measured using the Cambridge algorithm represents the limit to which
mis-clustering could be suppressed in the perturbative large
$y_\mathrm{cut}$ region.
  In this sense, the ansatz made earlier that the contamination
probability $p$ has a mainly nonperturbative origin is reasonable.

  On the other hand, towards very low $y_\mathrm{cut}$, as we have
observed above, the invariant mass, as in the JADE algorithm, may become a
better resolution variable than $k_T$. Hence the measurement of minijet
charge using these two algorithms could provide complementary information.

 \EPSFIGURE[ht]{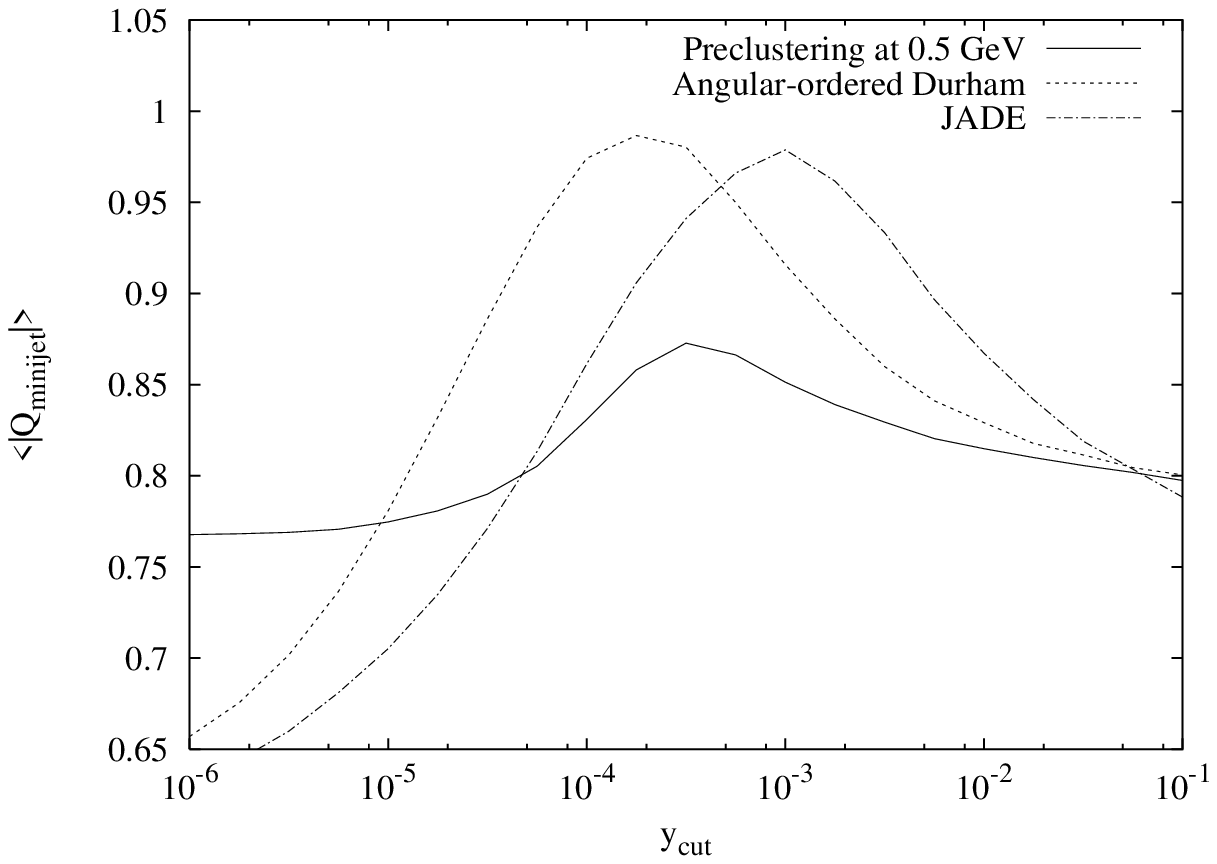,width=14cm}
 {Minijet charge in the 0.5 GeV `preclustered' angular-ordered Durham
algorithm compared with the JADE and angular-ordered Durham algorithms.
 \label{fig_tangle}}

  This suggests the introduction of a new class of scale-dependent jet
algorithms, which combines the low $y_{ij}$ behaviour of JADE with the
large $y_{ij}$ behaviour of $k_T$-based algorithms.
  As an example, we propose a `preclustered' scheme, where tracks are
first combined using the JADE algorithm up to some point given by the
resolution parameter $y_\mathrm{had}$, after which the tracks are combined
using a $k_T$-based algorithm with resolution parameter $y_\mathrm{cut}$.
  The case that uses the angular-ordered Durham algorithm is shown in
fig.~\ref{fig_tangle}. The preclustering cut-off $y_\mathrm{had}$ is the
normalization times (0.5 GeV)$^2$. This value is the most successful that
we have been able to find when using HERWIG with default parameter values.

  There is marked improvement, in terms of the peak height and large
$y_\mathrm{cut}$ behaviour, compared with either JADE or the
angular-ordered Durham algorithm. There is also improvement compared with
the Cambridge algorithm.

  We found that there is less marked improvement when preclustering is
followed by either the non-angular-ordered Durham algorithm or the
Cambridge algorithm, and there is no visible improvement when the JADE
preclustering phase is also angular-ordered.

  The preclustered curve remains large at low $y_\mathrm{cut}$ compared
with the JADE and angular-ordered Durham curves, and in the low
$y_\mathrm{cut}$ limit tends to the value obtained using the JADE
algorithm at the point $y_\mathrm{cut}=y_\mathrm{had}$.

  The sensitivity of the minijet charge to the parameters affecting
hadronization, as well as to the details of the jet algorithm, suggests
that because of our limited knowledge about the dynamics of hadronization,
the prediction of any existing Monte-Carlo event generator can not be
completely trusted when charge distribution is concerned. On the other
hand, the violation of local charge compensation during the hadronization
phase is presumably a universal phenomenon in all models of hadronization
that are based on preconfinement. Thus the general behaviour of the
minijet charge observable, that peaks around the hadronization region, is
a concrete prediction for this class of models.

 \EPSFIGURE[ht]{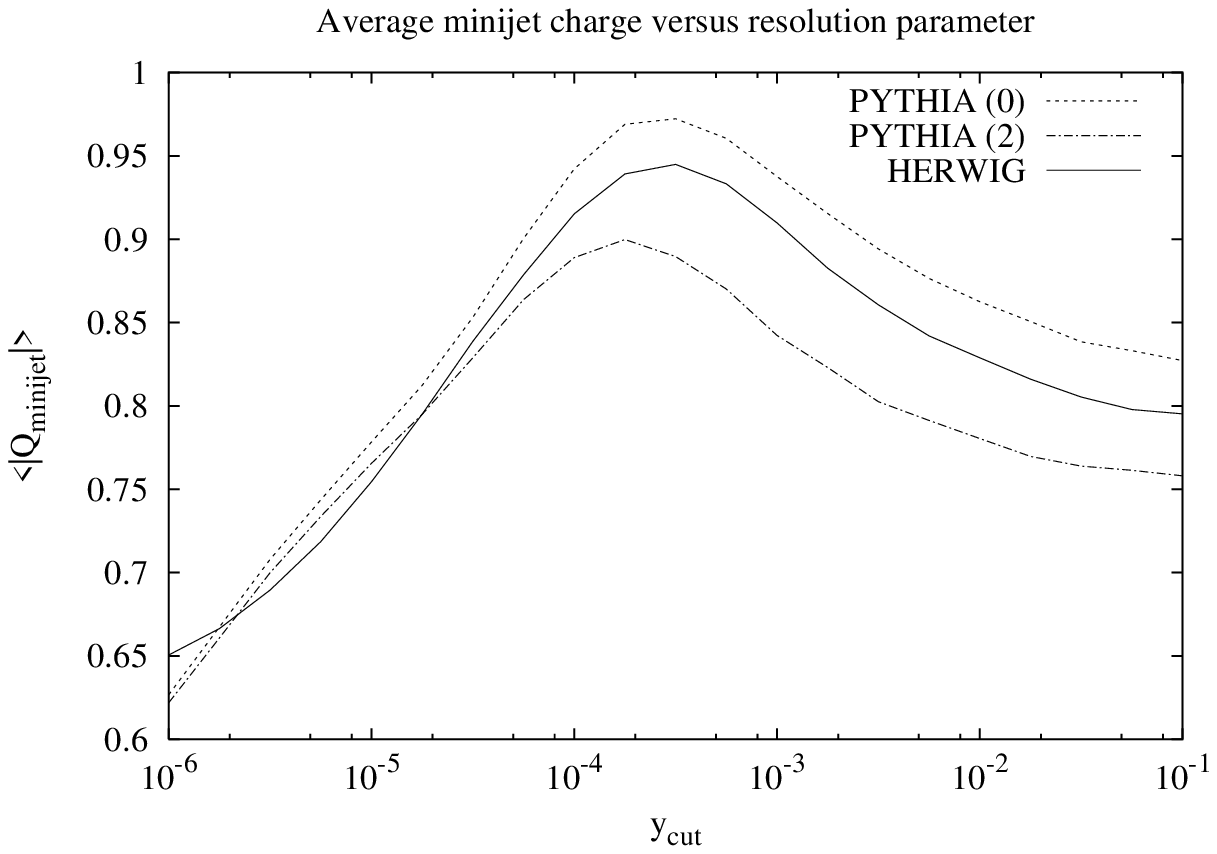,width=14cm}
 {A comparison of HERWIG and PYTHIA for the mean minijet charge. The lines
PYTHIA (0) and PYTHIA (2) correspond to the uncorrected option and the
second-order matrix-element corrected option respectively for the $Z^0$
decay in PYTHIA.
 \label{fig_pythia}}

  To demonstrate this point, in fig.~\ref{fig_pythia}, we show the numbers
obtained using PYTHIA. The overall behaviour is fairly similar, except the
behaviour at very small $y_\mathrm{cut}$, and there is an unwelcome
dependence on whether the $Z^0$ decay is matrix-element corrected,
although we have found that the difference between the first-order and
second-order corrected options is small. We have confirmed that the
corresponding difference between the matrix-element corrected and
uncorrected options is almost negligible in HERWIG. The effect of adding
the matrix-element correction in HERWIG is merely to raise the charge
slightly at larger values of $y_\mathrm{cut}$ as expected.

  As mentioned before, although the PYTHIA picture is that of string
fragmentation and the mass of the string may be large to start off with,
the scale at which charge is compensated is much smaller because the
fragmentation proceeds by the creation of quark-antiquark pairs. We see
from the figure that the peak position is roughly the same as that of
HERWIG.

 \EPSFIGURE[ht]{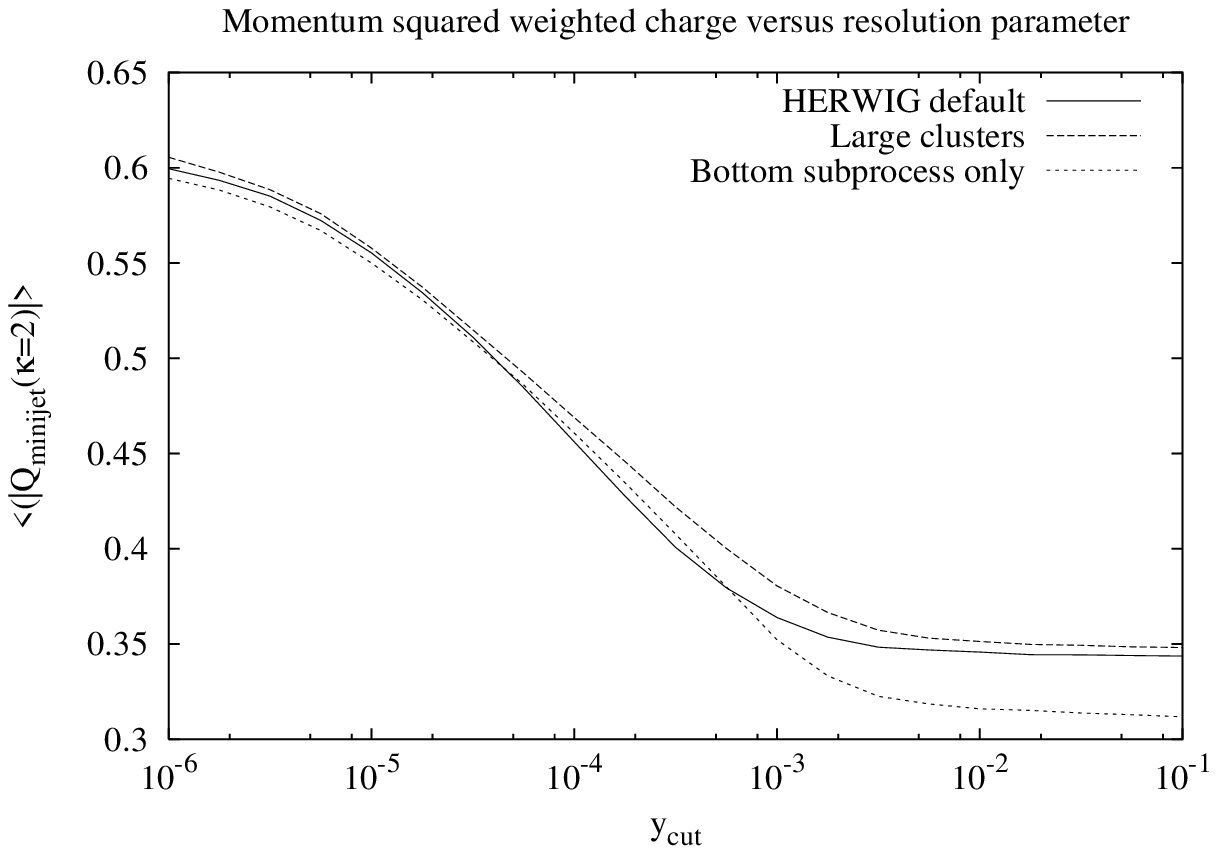,width=14cm}
 {The momentum-squared weighted minijet charge, plotted against the
Cambridge algorithm resolution parameter $y_\mathrm{cut}$. In addition to
the HERWIG default (solid line) and the large cluster (dashed line)
settings, we also show the bottom quark contribution by the dotted line.
 \label{fig_kappa2}}

  In fig.~\ref{fig_kappa2} we plot the momentum squared weighted minijet
charge, which corresponds to $\kappa=2$ in eqns.~(\ref{eqn_kappajet}) and
(\ref{eqn_kappajets}). The bottom quark contribution is plotted on a
separate curve in order to allow comparison with ref.~\cite{aleph_ab} in
the two-jet (large $y_\mathrm{cut}$) limit.

  The general behaviour is quite different from the unweighted charge of
fig.~\ref{fig_unweightedcharge}.
  In the low $y_\mathrm{cut}$ region, the weighted charge tends to the
same value as the unweighted charge. This is interpreted as the average
hadron charge.
  In the high $y_\mathrm{cut}$ region, the charge is much lower compared
with the unweighted charge. This is because the observable suffers less
from `contamination' due to soft tracks.

  We have also found that for a large and increasing $\kappa$ ($10\sim50$)  
and in the two-jet limit, the charge slowly increases in the region
$0.5\sim0.6$. In the limit of large $\kappa$, the momentum-weighted charge
tends to the charge of the highest energy hadron in each jet. From
figs.~\ref{fig_unweightedcharge} and \ref{fig_kappa2} we estimate the
average hadron charge to be slightly above $0.6$, in agreement with the 
above finding.

  For moderate values of $\kappa$, it is not reasonable to assume that
only the highest energy hadron contributes. In this case, the charge is
estimated as follows. Let us assume that the highest energy hadron
originates from the isotropic $n$-body decay of a heavier object.  If this
object is a cluster, $n$ is on average $\sim4$, whereas if it is a heavy
intermediate hadron, $n$ is typically 2. It is a reasonable approximation
to assume that this object has charge $0$ or $\pm1$, and it is also
reasonable to assume that the charge of each of the decay products is also
$0$ or $\pm1$. Then the expectation value for the momentum-weighted charge
of this object, when it decays into $\#_\mathrm{tot}$ objects out of which
$\#_\mathrm{ch}$ are charged, is simply:
 \begin{equation}
  <\!\mathrm{Q}(\kappa)\!>=\int_0^1
  \left(\prod_{i=1}^{\#_\mathrm{tot}}dz_i\right)
  \;\delta\!\left(\sum_{i=1}^{\#_\mathrm{tot}}z_i-1\right)
  \frac{
  \left|\sum_{i=1}^{\#_\mathrm{ch}}z_i^\kappa\left(-1\right)^i\right|}
  {\sum_{i=1}^{\#_\mathrm{tot}}z_i^\kappa}. \label{eqn_kappaobj}
 \end{equation}
  We assumed massless kinematics.
  For small $\#_\mathrm{tot}$ or $\#_\mathrm{ch}$, this 
multi-dimensional
integral can be evaluated analytically, but the general case is presumably
best evaluated numerically.


  Let us now adopt the viewpoint that the momentum-weighted charge is
mainly due to the heavy intermediate hadron which decays by 2-body
kinematics to give the highest energy hadron in the jet. The other
contributions are expected to mainly affect the denominator of
eqn.~(\ref{eqn_kappaobj}) by increasing it, as uncorrelated contributions
to the numerator cancel on average so long as they are not too large.

  For the $\#_\mathrm{tot}=2$ case and
$\kappa=2$, we obtain 0, 0.5 and $\log2=0.693$ respectively for
$\#_\mathrm{ch}=0,1,2$.
  Disregarding the contribution from other hadrons, we can estimate the
momentum-weighted charge as some weighted average of these three numbers.
  We note that out of the three numbers, only the $\#_\mathrm{tot}=1$ case
corresponds to nonzero overall charge. Hence one possible estimate would
be $0/4+0.5/2+0.693/4=0.42$.
  This is somewhat large compared with the large $y_\mathrm{cut}$ end of
fig.~\ref{fig_kappa2}, such that we see that the contribution from other
hadrons is not negligible but is under control.

  Our numbers for the mean momentum-weighted jet charge can be compared
with the hemisphere charge separation measurement in ref.~\cite{aleph_ab}.  
In the two-jet limit, the momentum component along the jet axis would not
be very different from the momentum component along the thrust axis, such
that we expect that the mean weighted charge presented here would
correspond to a half of the charge separation of ref.~\cite{aleph_ab}.
However, a quick comparison, for example with their fig.~2, shows that
there is nearly a factor of two difference. The hemisphere charge
separation is much smaller than is expected from HERWIG. This may be due
partly to the `detector effects' included therein. Furthermore, the
numbers due to JETSET presented in their tab.~4 has much greater flavour
dependence than in our work, even in the $\kappa\to\infty$ limit.
  Further study is desired in order to elucidate the nature of the
differences.

 \EPSFIGURE[ht]{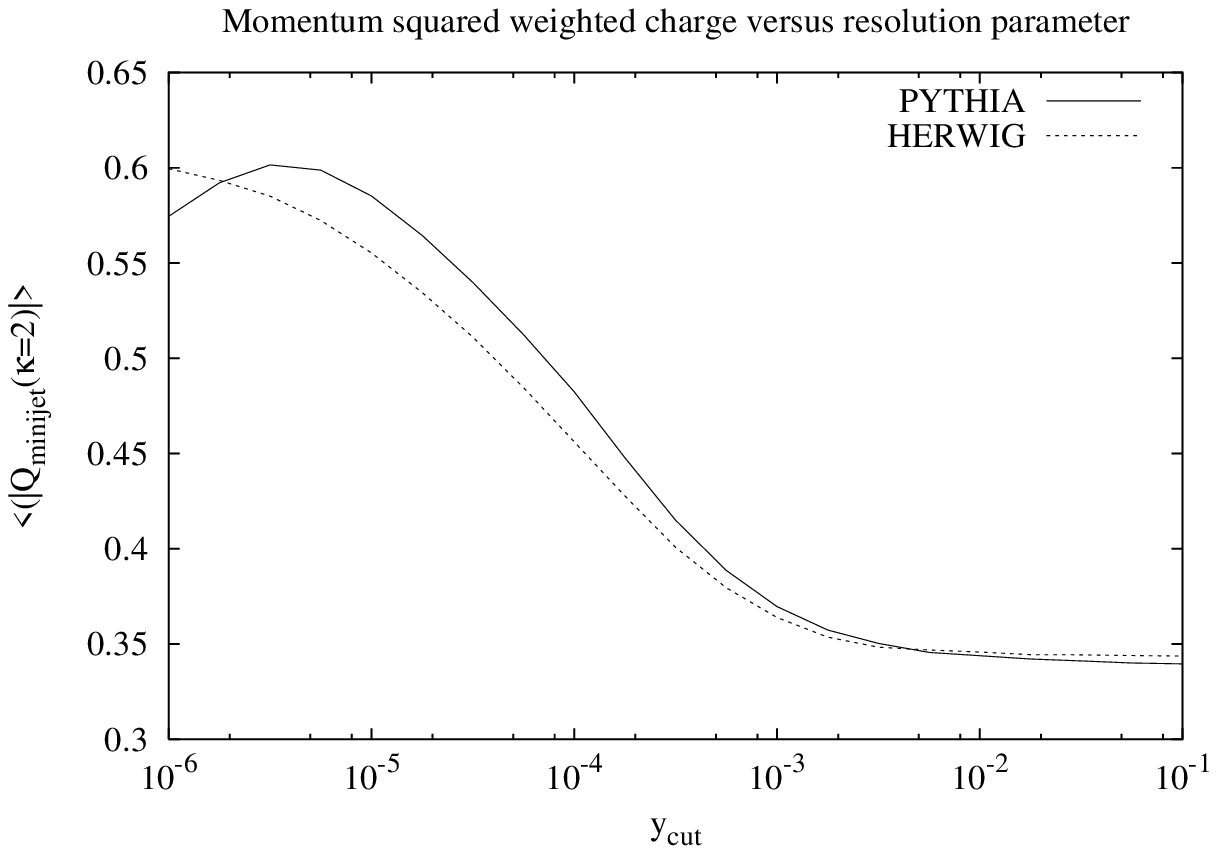,width=14cm}
 {A comparison of HERWIG and PYTHIA for the momentum-squared weighted 
minijet charge. The uncorrected $Z^0$ decay option is adopted for PYTHIA.
 \label{fig_pythia_kappa2}}

  For comparison, in fig.~\ref{fig_pythia_kappa2}, we show the result due
to PYTHIA (JETSET is now part of PYTHIA). In the two-jet limit, the two
generators give roughly equal numbers, such that the difference between
HERWIG and ref.~\cite{aleph_ab} is most likely not due to the difference
in the generator.
  For small $y_\mathrm{cut}$, the difference that is already visible in
fig.~\ref{fig_pythia} is magnified and the PYTHIA numbers show a peak at a
few times $10^{-6}$.

  The numbers shown so far have been obtained for the simulation of
$e^+e^-$ collisions at the $Z^0$ pole. On the other hand, considering the
high energy experiments that are currently in plan, it is interesting to
consider how our procedure and results are modified for collisions in a
hadron collider environment, where there are contaminations from emission
near the beam axis from the initial-state parton and also from the soft
underlying event.

  We generate QCD $2\to2$ hard scattering events in $p\bar p$ collision at
2 TeV in the central region. We require for both outgoing partons rapidity
$|y|<0.5$ and moderate transverse momentum 30 GeV $<p_T<$ 50 GeV. Events
are generated with and without soft underlying events simulated using the
HERWIG default procedure and with default parameters.

  Minijets are defined using the $k_T$-type jet algorithm of KTCLUS
\cite{ktclus} in this case as follows. We first define macrojets by
selecting a large enough $y_\mathrm{cut}$ value such that exactly two jets
remain uncombined with the beam. We then construct subjets constituting
these two macrojets at new values of $y_\mathrm{cut}$. These subjets are
regarded as the relevant minijets. In effect, we are studying the charge
structure of tracks within the two macrojets. The normalization factor in
eqn.~(\ref{eqn_yij}) for $y_\mathrm{cut}$ is taken to be $1/M_{Z^0}^2$
rather than $1/E_\mathrm{vis}^2$, such that the scale matches that of the
$e^+e^-$ simulations.

 \EPSFIGURE[ht]{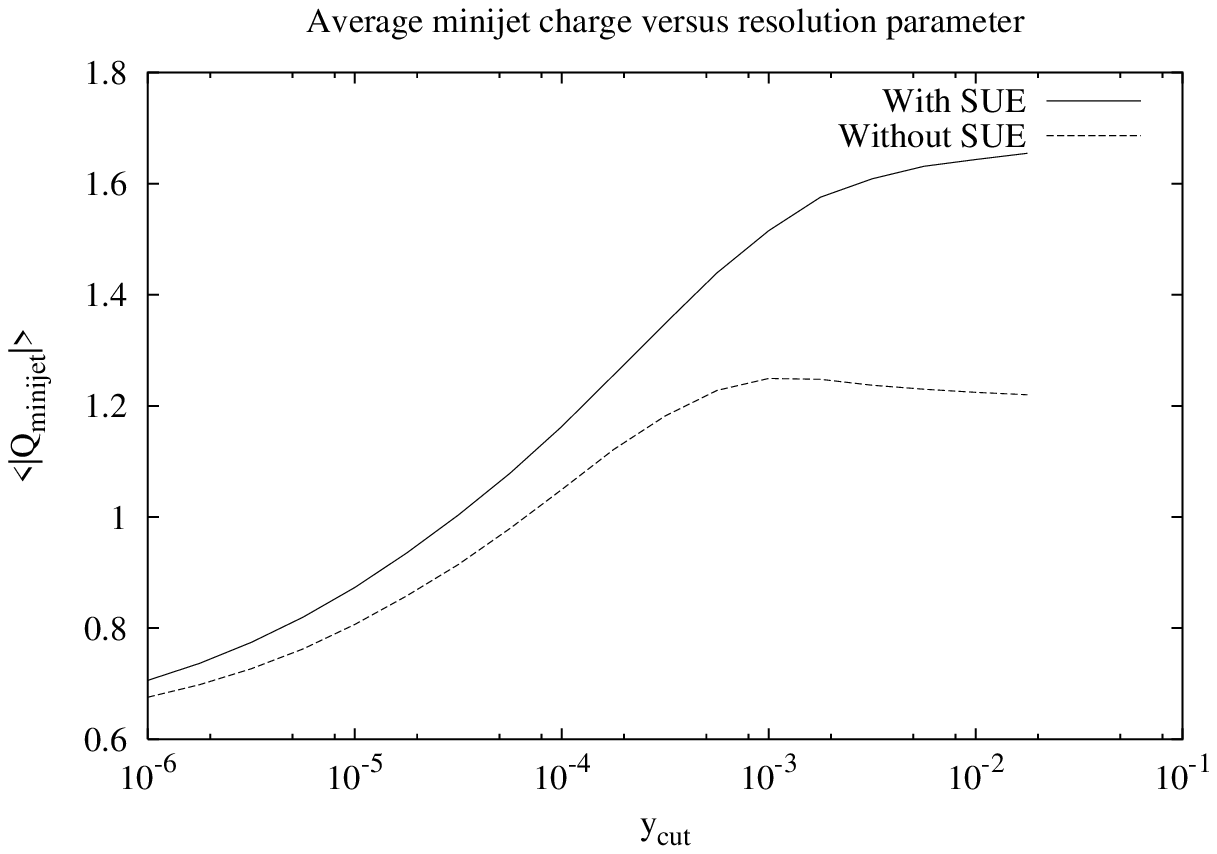,width=14cm}
 {Minijet charge in $p\bar p$ collision at 2 TeV. QCD $2\to2$ scattering
events are generated in the region $|y|<0.5$ and 30 GeV $<p_T<$ 50 GeV,
with and without soft underlying events. Only tracks belonging to the two
macrojets are considered. $y_\mathrm{cut}$ is normalized by $1/M_{Z^0}^2$.
 \label{fig_tev}}

  The result is shown in fig.~\ref{fig_tev}. As the jet algorithm of
KTCLUS is not angular-ordered, we should compare the result against the
Durham algorithm shown in fig.~\ref{fig_durham}. When the soft underlying
events are suppressed, the low $y_\mathrm{cut}$ behaviour is similar to
that of fig.~\ref{fig_durham}, but the peak has weakened and the minijet
charge remains large at large $y_\mathrm{cut}$.
  From this curve alone, it seems that the best that can be achieved in a
hadron collider environment is a fit with Monte Carlo simulations. On the
other hand, further refinement of the track selection procedure may
improve the prospect. The additional consideration of the JADE-type
algorithm is another possibility.

  The effect of adding soft underlying events is quite marked. However, we
note that the default procedure for soft underlying events in HERWIG
violates local charge compensation such that the numbers may be regarded
as a pessimistic estimate. In the alternative HERWIG-based simulations of
ref.~\cite{jimmy}, as well as in PYTHIA, the underlying events are
generated as multiple parton scattering and so charge is locally
compensated per each scattering. Having said this, whether charge is
locally compensated in soft underlying events at hadron colliders is an
open question which merits further investigation.

 \section{Interpretation in terms of cluster dynamics}
 \label{sec_interpretation}

  As mentioned in the previous sections, the mass of the perturbative
colour-preconfined clusters is of order of the cut-off scale. Put in other
words, emissions which are normally considered unresolvable in the parton
shower in fact affects the mass of clusters.
  If so, it is perhaps natural to consider the splitting of large
perturbative clusters, that occurs in HERWIG, in terms of an extension of
perturbative dynamics using a modified running $\alpha_S$, which describes
emission below the cut-off scale `before confinement'.

  There are several arguments in favour of this modification. First, in
the default HERWIG picture, the large clusters are split by a string-like
mechanism, by creating quark-antiquark pairs in the string vacuum. On the
other hand, the momentum direction of the quark and the antiquark
constituting the decaying cluster is conserved as if the cluster is still
an object composed of a perturbative quark and antiquark, unlike in the
subsequent cluster decay which is isotropic.
  Second, and in principle, this substitution could lift the scale
dependence in HERWIG between the parton shower phase and the cluster
splitting phase such that the ambiguity due to the incomplete description
of hadronization is in principle reduced to still lower scales.
  Third, the necessity to invoke an uncalculable string-like mechanism is
removed, and we would instead have a description in terms of a modified
$\alpha_S$, whose properties could be subjected to further study.

 \EPSFIGURE[ht]{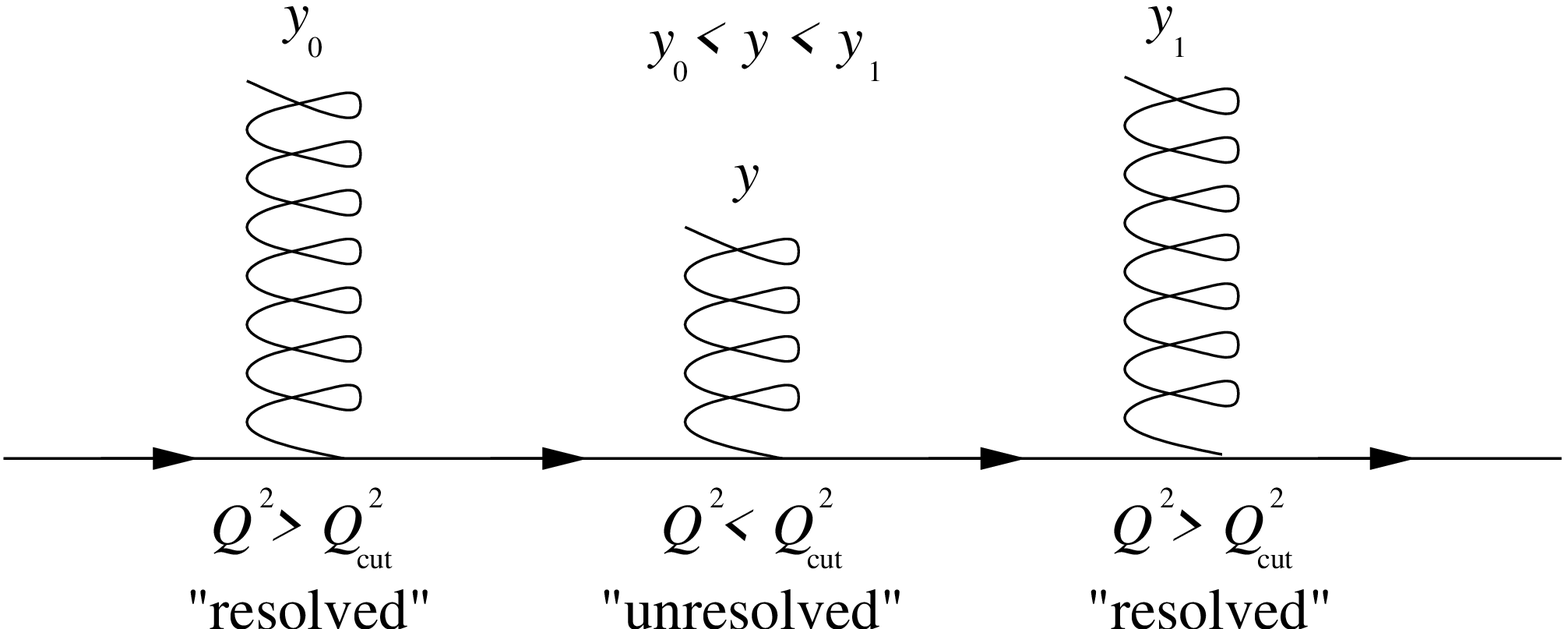, width=10cm}
 {The addition of an `unresolved' gluon to an angular-ordered parton
shower in between colour connected partons (gluons). $y$ is the
rapidity of the emitted parton.\label{fig_unresolved}}

  Before going into a discussion of the meaning of this $\alpha_S$, let us
first consider the Sudakov form factor due to the left-over `unresolved'
emissions. The addition of an `unresolved' gluon to an angular-ordered
parton shower is illustrated in fig.~\ref{fig_unresolved}, where we
consider the emission of an unresolved gluon in between rapidities $y_0$
and $y_1$, where $y_0$ and $y_1$ are the rapidities of the gluons which
have been emitted in the course of the resolved parton shower.

  The emission probability of a gluon in a phase space element
$(dQ^2,dy)$, where $Q^2$ and $y$ are the $p_T^2$ and rapidity of the
emitted gluon respectively, in the soft limit, is:
 \begin{equation}
  dP_\mathrm{emission}=
  \frac{dQ^2}{Q^2}dy\frac{\alpha_S(Q^2)C}{\pi}
 \end{equation}
  The colour factor $C$ is $C_F$ for emission from a quark and $C_A$ for
emission from a gluon.
  As we are considering cluster dynamics and the gluon is effectively a
quark--antiquark object in this respect, the colour factor is $C_F$, or
equally well $C_A/2$, at leading order in $1/N_C$.

  The equivalence of this expression with the ordinary angular-ordered
parton shower \cite{coherent_emission} in the soft limit is demonstrated
as follows. Let us consider the emission probability from a quark in the
splitting $q_a\to q_bg_c$:
 \begin{equation}
  dP_\mathrm{emission}(q_a\to q_bg_c)=
  \frac{d\tilde t}{\tilde t}\frac{dz_b}{2\pi}\alpha_S(Q^2)
  \widehat P_{qq}(z_b).
 \end{equation}
  Here $\tilde t=E_a^2(1-\cos\theta_{bc})$. $E_a$ is constant until an
emission takes place, such that $d\tilde t/\tilde
t=d\theta^2/\theta^2=2dy$ in the soft and the usual small angle limit. In
the soft limit, the splitting function $\widehat P_{qq}(z_b)=2C_F/z_c$,
such that we have:
 \begin{equation}
  dP_\mathrm{emission}(q_a\to q_bg_c)=
  2dy\frac{dz_b}{2\pi}\alpha_S(Q^2)\frac{2C_F}{z_c}.
 \end{equation}
  Now replacing $dz_b$ by $dz_c$ and using $Q^2=2\tilde tz_c^2$ in the
soft limit, we see that the two expressions are equivalent.

  For an angular-ordered parton shower, the emission is ordered in $y$.
  The Sudakov form factor is:
 \begin{equation}
  \Delta_q(y)=\exp\left[-\int\frac{dQ^2}{Q^2}dy
  \frac{\alpha_S(Q^2)C_F}{\pi} \right].
 \end{equation}
  We are only interested in emission below the cut-off scale, such that
the $Q^2$ integration factorizes out:
 \begin{equation}
  \Delta_q(y)=\left[\exp(y-y_0)\right]^{-I_0},\label{eqn_sudy}
 \end{equation}
  where
 \begin{equation}
  I_0 = \int_0^{Q^2_\mathrm{cut}}
  \frac{dQ^2}{Q^2}\frac{\alpha_S(Q^2)C_F}{\pi}.
 \end{equation}
  The integral is finite if $\alpha_S$ tends to zero as a power as $Q^2$
tends to zero. A possible physical argument would be to say that if colour
is confined in the large distance limit,
  then any $\alpha_S$ with some physical interpretation as the coupling of
coloured objects can be defined to vanish as a power.

  We may thence carry on to estimating $I_0$ in the above using models of
low energy $\alpha_S$. As an example, and not necessarily an appropriate
one, we may consider the effective $\alpha_S$ of
ref.~\cite{webber_alphas}, which is derived from the general consideration
of power-suppressed corrections to event-shape observables. In this
approach, the renormalon chain which gives rise to the power-suppressed
corrections is summed using a dispersion relation. The effective
$\alpha_S$ is defined to be analytic everywhere except along the negative
real $Q^2$ axis, and its infrared behaviour is connected to the power
corrections through the analyticity of the observable under attention, or
more precisely its characteristic function, as a function of small squared
`gluon mass'.

  One difficulty when using this $\alpha_S$ to estimate $I_0$ is that it
remains finite as $Q^2\to0$ as shown in fig.~\ref{fig_alphas}, such that
its integral is not finite. In the dispersive resummation of the
renormalon chain, this is not a problem as only the moments of $\alpha_S$
are related to the power-suppressed corrections.
  For now we content ourselves by cutting off at $Q\sim0.1$ GeV and
integrating up to the HERWIG cut-off at around {\tt RMASS(13)}=0.75 GeV.
We obtain $I_0\sim 1$.

  We shall discuss the above points regarding the behaviour of low-energy
$\alpha_S$ in more detail in sec.~\ref{sec_discussion}.

 \EPSFIGURE[ht]{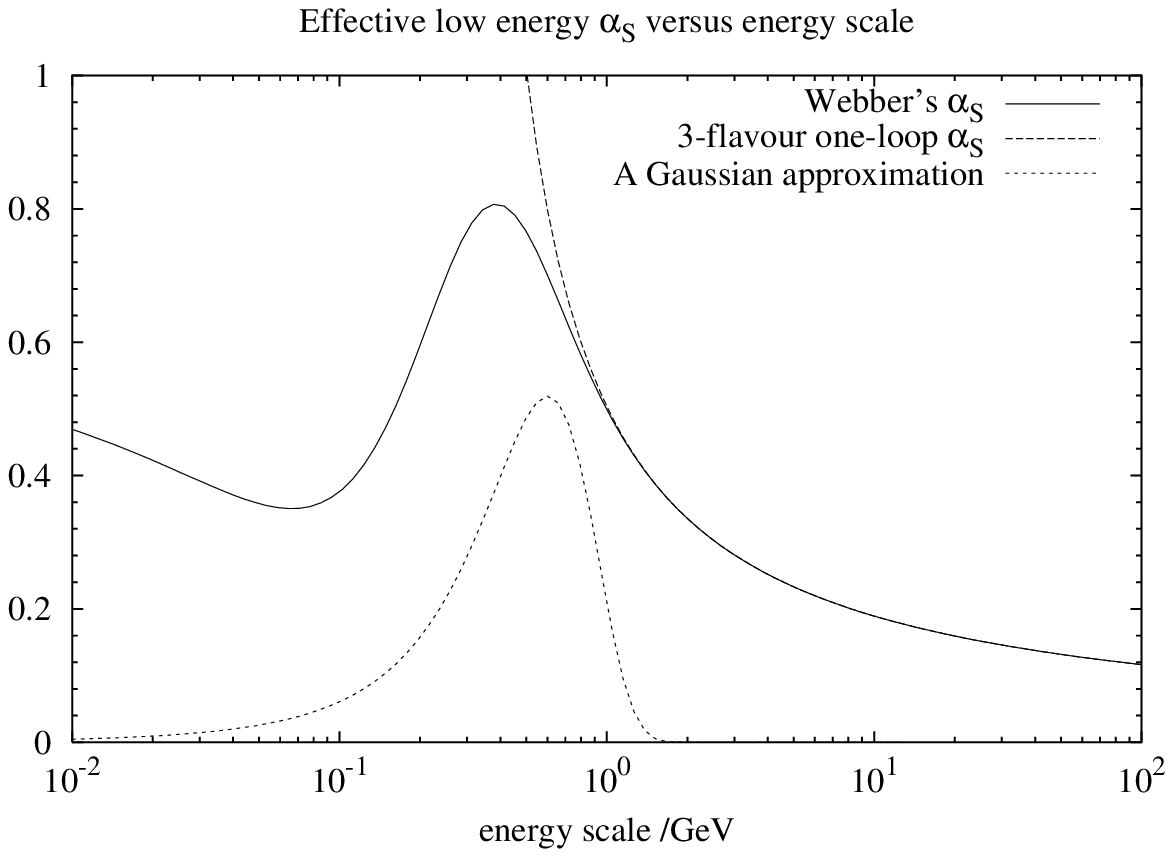, width=14cm}
 {The effective $\alpha_S$ of ref.~\cite{webber_alphas}, derived from the
general consideration of power-suppressed corrections to event-shape
observables.
  The effective $\alpha_S$ coincides with the three-flavour $\alpha_S$ in
the high-energy limit. The three-flavour $\Lambda_\mathrm{QCD}$ is taken
to be 0.25 GeV.
  We also show a Gaussian approximation, with arbitrary normalization.
 \label{fig_alphas}}

  It is a simple matter to relate the emission dynamics to the kinematics
of cluster splitting. Let us consider the splitting of a dipole with mass
$M_0$ into two smaller dipoles with masses $M_1$ and $M_2$. The kinematics
has already been calculated in the context of the dipole cascade formalism
in ref.~\cite{ariadne}.
  As the integral is Lorentz invariant, let us transform to the cluster
frame. Neglecting quark masses, rapidity in this case is:
 \begin{eqnarray}
  y &=& \log(M_2/M_1),\\
  |y|&\lesssim& \log(M_0/Q_\mathrm{cut}).\label{rapidity_limits}
 \end{eqnarray}
  Combining these with eqn.~(\ref{eqn_sudy}) and $I_0\sim1$, we may
estimate the probability of there being no splitting. If the cluster mass
is ten times greater than the cut-off, for example, the probability of the
cluster surviving without emission is only 1\%.

  In general, as can be inferred from eqns.~(\ref{eqn_sudy}) and 
(\ref{rapidity_limits}), $I_0$ governs the average number of smaller 
clusters that a large cluster splits into, viz.:
 \begin{equation}
  <\!\#_\mathrm{split}\!>\sim I_0\Delta y\approx
  I_0\log\frac{M_0^2}{Q^2_\mathrm{cut}}.\label{eqn_splitmulti}
 \end{equation}
  From fig.~\ref{fig_clusmas}, we can estimate the value of
$<\!\!\#_\mathrm{split}\!\!>$ that leads to the correct cluster
multiplicity. We estimate this number to be $\sim3$ and $M_0$ to be $\sim$
10 GeV. This gives $I_0\sim0.5$ and thus $I_0\sim1$ obtained above is too
large.

  $Q^2=p_T^2$ on the other hand controls the typical cluster mass, and can 
be expressed as:
 \begin{equation}
  Q^2 =\frac{M_1^2M_2^2}{M_0^2}.\label{dipole_scale}
 \end{equation}
  If we omit the consideration of the nonperturbative $g\to q\bar q$
splitting that converts the dipoles into clusters, the physics of which we
do not yet understand and which we will discuss in
sec.~\ref{sec_discussion}, the cluster splitting is henceforth governed by
the above equations.
  One should add a technical detail, that in HERWIG, the cluster mass that
is considered has the constituent quark masses subtracted.

  We note that the default cluster splitting procedure of HERWIG violates
the above expression, eqn.~(\ref{dipole_scale}), for $Q^2$, when the
product of the masses of the two decay products divided by the decaying
cluster mass is greater than the parton shower cut-off.

  Based on the above considerations, we carry on to creating a simple code
which modifies HERWIG to perform cluster decay by the modified-$\alpha_S$
approach. However, there are two subtleties that need to be dealt with.

  First, if there is a cascade splitting of clusters, one must be careful
not to double-count soft emission. The simplest method for avoiding
double-counting would be to introduce a flag.
  On the other hand, when making a HERWIG implementation, as there is a
cut-off on the maximum allowed cluster mass, this problem is not severe.

  Let us consider again the cluster mass distribution shown in
fig.~\ref{fig_clusmas}. The parton shower cut-off, which also signifies
the scale of the low-energy cluster-splitting dynamics, is of order {\tt
RMASS(13)}=0.75 GeV, whereas the minimum cluster mass that is split is
around {\tt CLMAX}=3.35 GeV. We see that in most of the cases, provided
that eqn.~(\ref{dipole_scale}) is satisfied, at least one of the decay
products is lighter than the cut-off. Hence there would be no further
emission from this lighter cluster, and by adopting an ordering that
starts from the largest value of $|y|$, the problem would be resolved.

  Second, and related to the first point, the ordering in $y$ is not
strictly necessary. So long as the cluster mass can be considered large
compared with the parton shower cut-off, the order in which the
`unresolved' gluons are emitted, in principle, does not matter.
  However, in a practical HERWIG implementation, because of the cut-off on
the maximum allowed cluster mass and the incomplete description of physics
below this scale, there does remain some dependence on the ordering.

  In particular, when using the above effective $\alpha_S$, because
$I_0\sim1$ is large, we have found that if emission is in the order of
decreasing $|y|$ as suggested above, too many clusters are formed with
small mass. This is unwelcome in the sense that these small clusters can
not be considered perturbative objects. On the other hand, when
$I_0\sim0.5$ as is preferred from multiplicity considerations, we have
found that there is not much dependence on whether $|y|$ is ordered. In
this case there is not a clear-cut connection between the presence of this
ordering and, for example, the cluster mass distribution, such that for
now, we may abondon the ordering in $y$ and generate it uniformly.

  After the above considerations, we implement the following procedure,
which is not precisely the procedure dictated by the modified-$\alpha_S$
approach, but more close to the HERWIG default cluster-splitting
algorithm:
 \begin{enumerate}

 \item Terminate the parton shower by means of an artificial gluon mass
{\tt RMASS(13)}=0.75 GeV, as according to the HERWIG default prescription.

 \item For clusters whose mass is greater than about {\tt CLMAX}, or more
precisely for a cluster with mass $M$ made from quarks $i$ and $j$:
 \begin{equation}
 M^\mathtt{CLPOW} > \mathtt{CLMAX}^\mathtt{CLPOW}+
 (\mathtt{RMASS}(i)+\mathtt{RMASS}(j))^\mathtt{CLPOW},\label{eqn_clpow}
 \end{equation}
  split the cluster, as follows.
 \label{step_split}

 \item Rapidity is first generated, not by following
eqn.~(\ref{eqn_sudy}), but as a flat distribution in the range suggested
by eqn.~(\ref{rapidity_limits}). This is correct in the limit where there
is one and only one `unresolved' emission. It is also correct in the case
where emission is ordered in $Q^2$, as in ref.~\cite{ariadne}. In this
case, the distribution of $Q^2$ described below should be interpreted as a
Sudakov form factor rather than $\alpha_S$.

 \item $Q^2$ is generated in between the parton shower cut-off scale and 
the lower cut-off scale for $\alpha_S$, according to the distribution:
 \begin{equation}
  \propto\frac{dQ^2}{Q^2}\alpha_S(Q^2).
 \end{equation}

 \item The process is repeated until there remains no more cluster that
satisfies eqn.~(\ref{eqn_clpow}).
  Thus all clusters above the cluster-mass cut-off are split. The error
due to disallowing some heavy clusters is negligible so long as the
probability of splitting, or the parameter $I_0$, is sufficiently large.

 \end{enumerate}

  In the above implementation, the normalization of $\alpha_S$, which
should affect multiplicity, is nevertheless irrelevant so long as we
abandon $y$ ordering.
  Hence one may as well substitute for low energy $\alpha_S$ by a Gaussian
distribution. A useful substitution may be:
 \begin{equation}
  \propto Q
  \exp\left[\frac{(Q-Q_\mathrm{mean})^2}{2(\Delta Q)^2}\right],
  \label{eqn_gaussian}
 \end{equation}
  such that $Q$ is generated with mean $Q_\mathrm{mean}$ and standard
deviation $\Delta Q$.
  Although this is a fairly economical parametrization, we may further
reduce the number of parameters by setting the upper limit equal to the
parton-shower cut-off. Thus we set:
 \begin{eqnarray}
  Q_\mathrm{mean}&=&\frac{\mathtt{RMASS}(13)+Q_\mathrm{min}}2,
  \label{eqn_qmean}\\
  \Delta Q       &=&\frac{\mathtt{RMASS}(13)-Q_\mathrm{min}}2.
 \end{eqnarray}
  The Gaussian distribution corresponding to $Q_\mathrm{min}=0.1$ GeV is
shown in fig.~\ref{fig_alphas}.

  The cluster mass distribution due to this choice is shown in
fig.~\ref{fig_modclusmas}, where we compare it against the HERWIG default
distribution and against the result of adopting the procedure above with
the $\alpha_S$ of ref.~\cite{webber_alphas}. When using this $\alpha_S$,
$Q^2$ is generated in between 0.1 GeV and 0.75 GeV.

 \EPSFIGURE[ht]{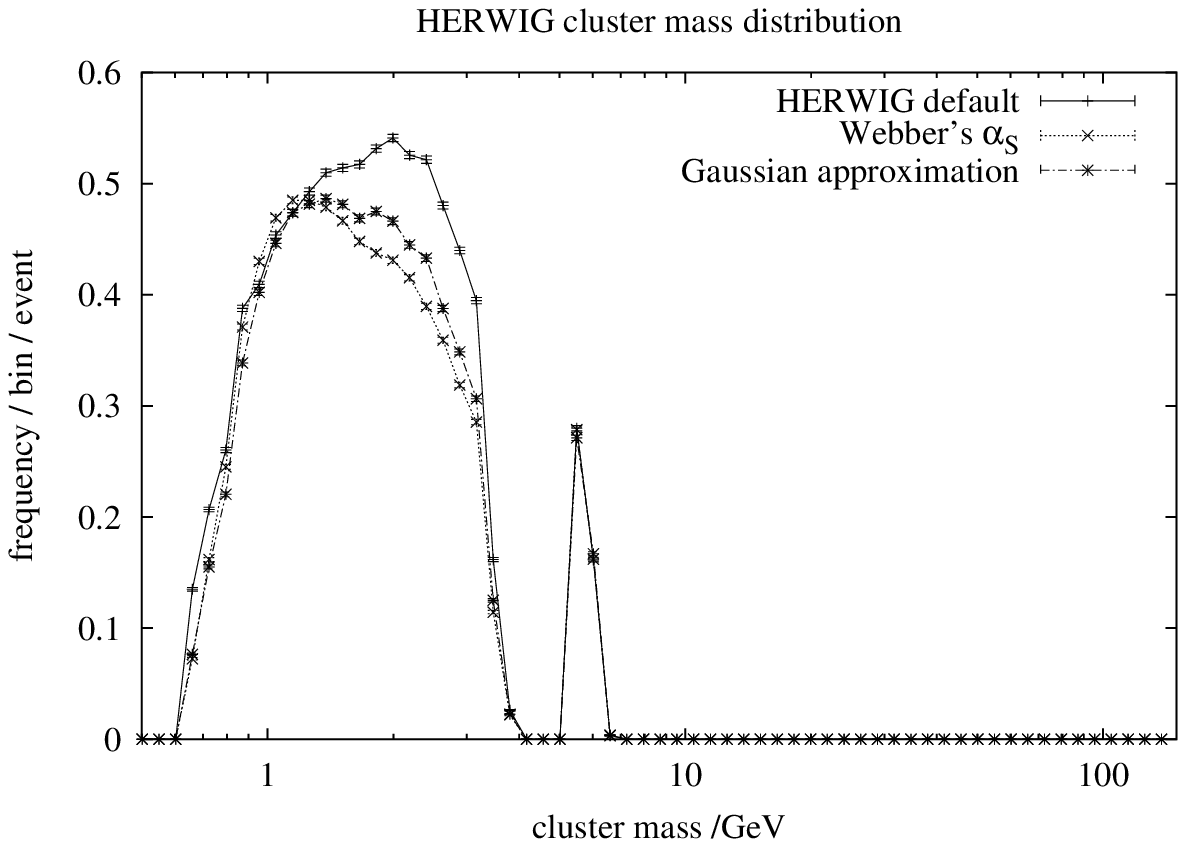, width=14cm}
 {The cluster mass distribution generated using the HERWIG default
procedure, the $\alpha_S$ of ref.~\cite{webber_alphas} and a Gaussian
approximation with $Q_\mathrm{min}=0.1$ GeV.
 \label{fig_modclusmas}}

  We see that with $Q_\mathrm{min}=0.1$ GeV, although the distribution is
close to the HERWIG default, the distribution is too skewed to the low
mass end, and multiplicity is not large enough. These two problems are
mutually related, but presumably the more serious is the skew to the
low-mass end, as the multiplicity is in principle affected by the
normalization of $\alpha_S$ as noted above.

  Both problems can be rectified by raising $Q_\mathrm{min}$, and we have
found that the correct enhancement of the high mass end is obtained at
$Q_\mathrm{min}\sim0.4$ GeV. The individual identified particle yields are
also affected by this cluster mass distribution. As the shape of the
distribution is not identical with the HERWIG default, the yields would be
slightly different, but as there is no reason to believe that the Gaussian
distribution resembles reality, there would be no reason to believe that
the resulting yields would be closer or farther to reality than the
default HERWIG numbers.

 \EPSFIGURE[ht]{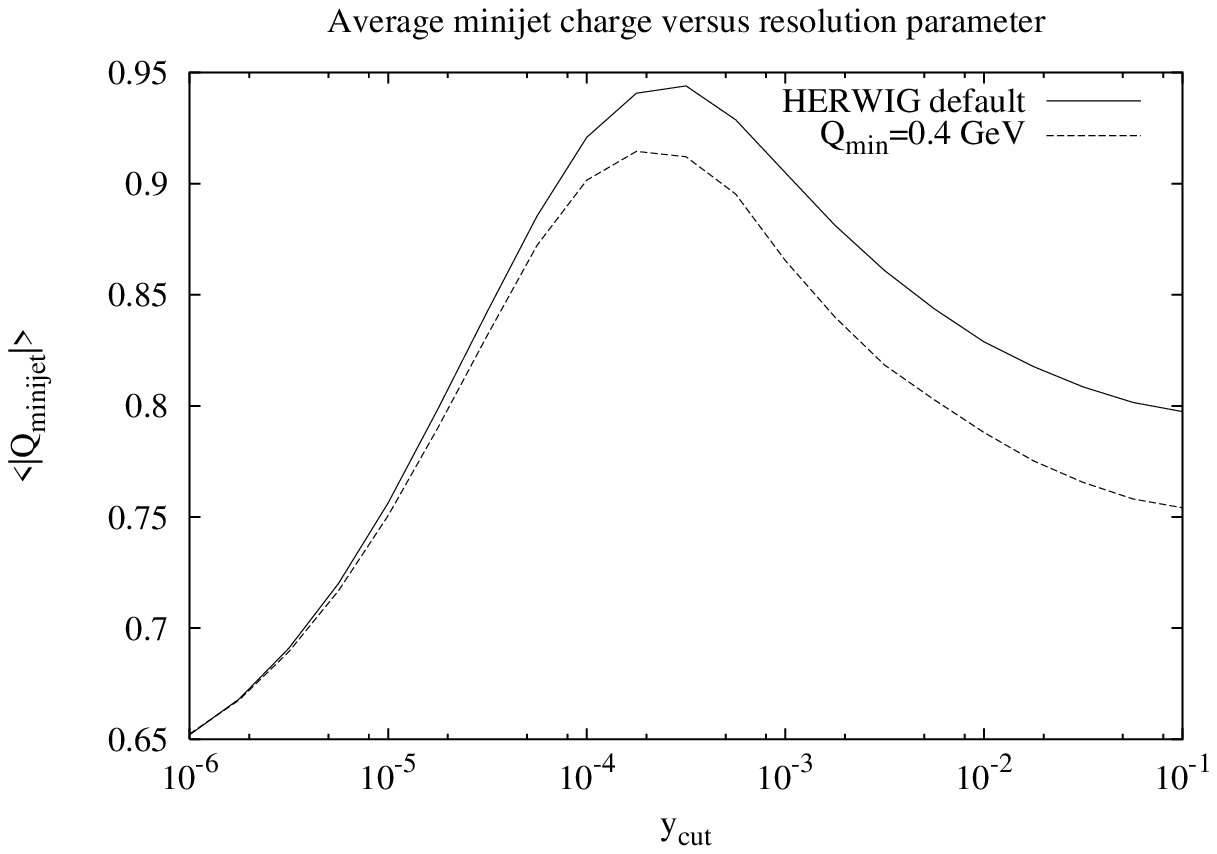,width=14cm}
 {The mean minijet charge plotted against the Cambridge algorithm
resolution parameter $y_\mathrm{cut}$.
  The sample size is 50000 events for each curve.
 \label{fig_gausscharge}}

  In fig.~\ref{fig_gausscharge}, we show the average minijet charge. The
numbers are stable with respect to small perturbation of the parameter
$Q_\mathrm{min}$. The result shows that the peak position in
$y_\mathrm{cut}$ has shifted slightly to the left compared to the HERWIG
default numbers, and the peak height is smaller. As the cluster mass
distribution itself is more or less unchanged, this modification should be
interpreted as the result of having less probability of a large cluster
splitting into two adjacent large clusters. Thus a modification of HERWIG
that excludes cluster splitting that can be interpreted as due to high
$p_T$ emission leads to the suppression of minijet charge.

  This argument also helps to understand why the PYTHIA minijet charge,
without the matrix element correction, is larger than the HERWIG result,
as seen in fig.~\ref{fig_pythia}.


  It is interesting to see whether event shape observables are also
affected. We have computed sphericity and found that the modification due
to the implementation of the new algorithm is slight.

 \section{Interpretation of low energy $\alpha_S$}\label{sec_discussion}

  What we have learnt from the above simulation in terms of the dynamics
of hadronization is that the scale relevant to the splitting of clusters
is between about 0.4 GeV and 0.75 GeV. In terms of the perturbative
$\alpha_S$, between about 0.5 and 1.

  What remains is to understand the origin of this structure. If we reduce
the scale of splitting, the cluster mass distribution is too skewed
towards the low mass end and multiplicity becomes insufficient. It would
not be sufficient to propose, for instance, that only the highest $p_T$
emission affects the splitting of clusters.

  Let us consider a toy model, in which perturbative emissions have
Sudakov form factor $\Delta(Q^2)$ and the confinement interaction has
Sudakov form factor $\Delta_C(Q^2)$. Then the probability of emission
`before confinement' is given by exponentiating the emission probability
times the survival probability and hence the following modified Sudakov
form factor:
 \begin{eqnarray}
  \Delta_\mathrm{effective}(Q^2) &=&
  \exp\left[\int \frac{dQ^2}{Q^2}
  \Delta_C(Q^2)\frac{d\log\Delta(Q^2)}{dQ^2}
  \right]\\
  &=&
  \exp\left[-\int \frac{dQ^2}{Q^2}dy
  \Delta_C(Q^2)\frac{\alpha_S(Q^2)C_F}\pi
  \right].
 \end{eqnarray}
  Although the confining potential is not obtained at any finite order of
perturbation theory, for the sake of discussion we may write confinement
as an $\mathcal{O}(\alpha_S^2)$ `gluon exchange' process. Thus
$\Delta_C(Q^2)=1-\mathcal{O}(\alpha_S^2)$ and hence we have a plausible
explanation for the suppression of emission at lower scales where
$\alpha_S(Q^2)$ becomes large.

  In terms of probability, this is equivalent to saying that although the
emission of gluons at low $Q^2$ has large probability, in the time that it
takes to have this emission, there is greater probability that the cluster
will be confined. Hence emission is suppressed.

  In terms of effective $\alpha_S$, the effective $\alpha_S$ that controls
`unresolved' emission that splits the cluster is expanded in terms of
perturbative $\alpha_S$ as $\alpha_S-\mathcal{O}(\alpha_S^3)$. However,
the coefficient may depend on the cluster mass and the nature of the
observable and hence this $\alpha_S$ is not necessarily universal.

  In the above model, we have assumed that the onset of confinement is
sudden. This is also the case in HERWIG, where the `confined',
`nonperturbative' clusters thus formed then decay isotropically.
  In reality, the soft gluon exchange interaction between the quark and
the antiquark in a cluster is expected to enter earlier. Hence clusters
would be `rotated' as illustrated in fig.~\ref{fig_confinement}. We expect
that this leads to the broadening of the peak in the minijet charge
observable, although a quantitative prediction is beyond the scope of this
study.

 \EPSFIGURE[ht]{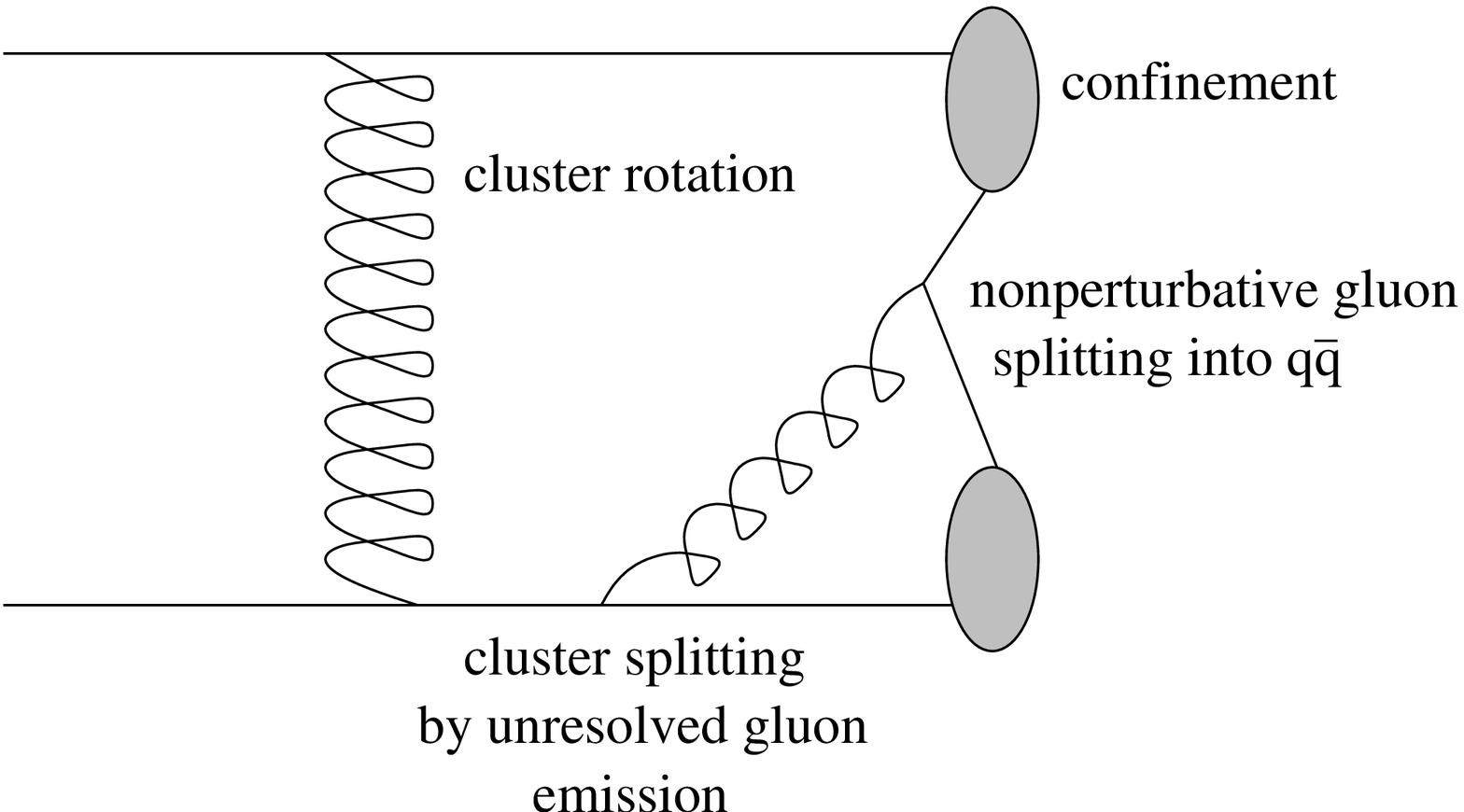,width=12cm}
 {Perturbative and nonperturbative dynamics affecting perturbative
clusters.
 \label{fig_confinement}}


  Let us illustrate and concretize the above discussion of `emission
before confinement' by considering the dynamics of the $g\to q\bar q$
splitting intrinsic to HERWIG which is not understood. If this splitting
is physical, for instance if confinement enhances this splitting, it would
be possible to define the corresponding beta function.
  Solving the renormalization group equation, we may define another
$\alpha_S$ that may be regular in the infrared.
  This provides another viewpoint that is compatible with the above toy
model in the sense that we may loosely identify `confinement' with the
$g\to q\bar q$ splitting.

  We may define the effective strength for gluon splitting, that could be
used to replace HERWIG's non-perturbative gluon-splitting, as follows:
 \begin{equation}
  T_\mathrm{eff}(Q^2)=\frac{11C_A}4-3\pi\beta(Q^2),
 \end{equation}
  where $\beta=d\alpha_S^{-1}/d\log Q^2$. $T_\mathrm{eff}=n_fT_R$ in
perturbation theory at the lowest order, but it becomes large at low
scales for any parametrization of $\alpha_S$ that turns over. If we
require that $\alpha_S$ vanishes as a power of $Q^2$ such that its
integral with respect to $\log Q^2$ is finite, we see that
$T_\mathrm{eff}$ must diverge towards $Q^2\to0$ with the same power, i.e.,
$\alpha_ST_\mathrm{eff}$ tends to a constant.
  Although this power behaviour can not be obtained at any finite order in
perturbation theory, it is precisely what one would expect as the form of
a nonperturbative higher-twist contribution. We may write:
 \begin{equation}
  \beta(Q^2) = \beta_\mathrm{pert.}(\alpha_S) - \frac{A}{Q^n}.
  \label{eqn_nonperturbative_beta}
 \end{equation}
 Dropping the higher order perturbative contributions, we obtain
$\alpha_S$ as:
 \begin{equation}
  \alpha_S(Q^2) = \frac{1}{\beta_0\log(Q^2/\Lambda^2)+2A/(nQ^n)}.
  \label{eqn_nonperturbative_alphas}
 \end{equation}
  Here $\beta_0=(33-2n_f)/12\pi$. Although the forced $g\to q\bar q$
splitting in HERWIG only involves two flavours, for the sake of
consistency with the previous discussions, let us adopt $n_f=3$ for now.
  This $\alpha_S$ has the desired properties of integrability under $\log
Q^2$ and remaining well-defined on the positive real axis, provided that
$A>\beta_0\Lambda^ne^{-1}$ such that the denominator is positive definite.
  On the other hand, when either the denominator is positive definite or
$n>4$, the expression has complex poles.

  If we require that this $\alpha_S$ is universal, we should require that
$n$ is large such that we do not introduce power corrections that are not
predicted by the OPE formalism to quantities that are proportional to
$\alpha_S(Q^2)$ in leading order such as the $e^+e^-$ total cross section.
OPE predicts a term proportional to the dimension-4 gluon condensate,
which is therefore $\propto 1/Q^4$ by dimensional analysis, whereas the
above $\alpha_S$ gives rise to terms $\propto(\alpha_S^\mathrm{PT})^2/Q^n$
for large $Q$, where $\alpha_S^\mathrm{PT}$ is the perturbative
$\alpha_S$, such that the $n=2$ case is forbidden.

  In the discussion of ref.~\cite{dispersive}, even additional
contributions proportional to $1/Q^4$ is disfavoured, as this gives an
additional `nonperturbative' contribution which is not related to the
region of small momentum flow, i.e. the region which is responsible for
the condensates that give rise to the power corrections, in the
corresponding Feynman diagrams.

 \EPSFIGURE[ht]{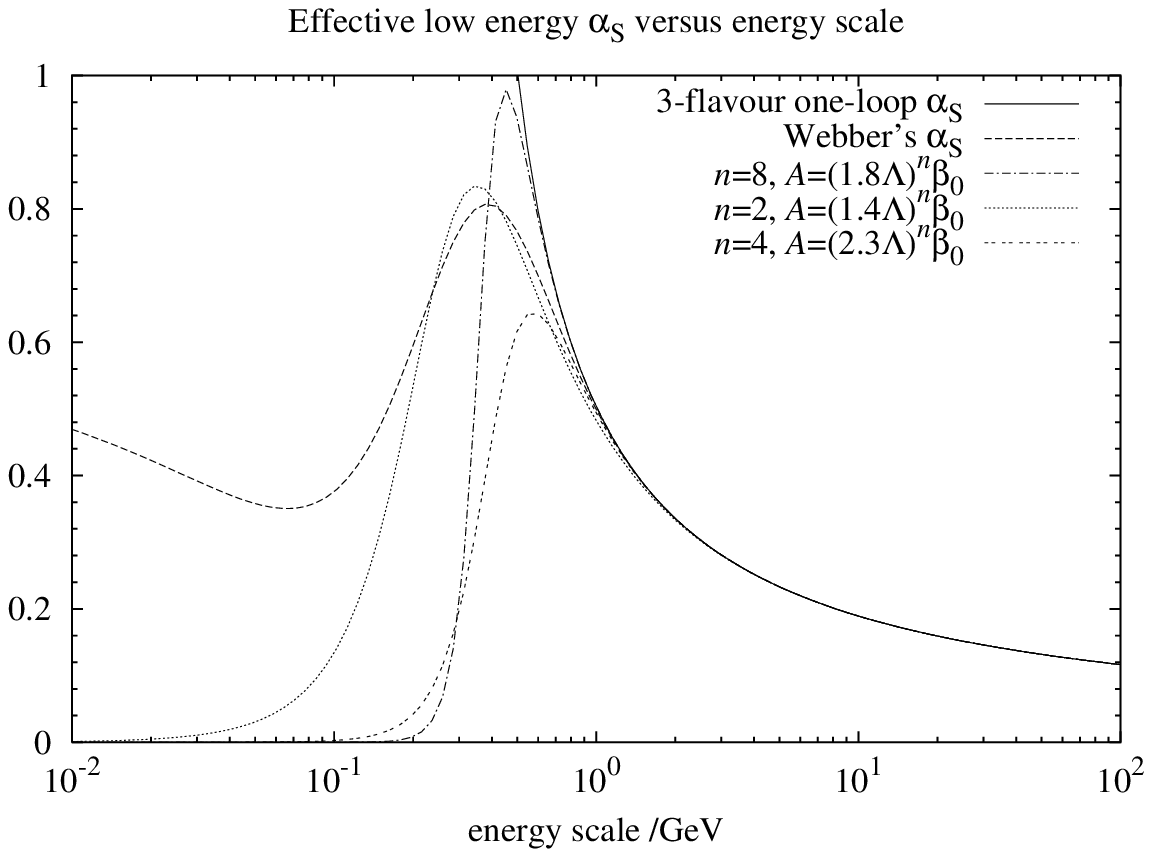, width=14cm}
 {The $\alpha_S$ derived from a power corrected beta function.
 \label{fig_alphas2}}

  As an example, let us adopt $n=8$ here, which is the same as the large
$Q^2$ behaviour of the effective $\alpha_S$ of ref.~\cite{webber_alphas}.
We set $A=(1.8\Lambda)^n\beta_0$ and display the behaviour of this
perticular choice in fig.~\ref{fig_alphas2}. For reasonable choices of the
parameter $A$, the low energy cut-off is large, in quantitative agreement
with our earlier finding that $Q_\mathrm{min}\sim0.4$.
  On the other hand, the low $Q^2$ behaviour of the aforementioned
$\alpha_S$ is matched better for small $n$. In fig.~\ref{fig_alphas2}, we
also plot the $n=2$ case for comparison, with $A=(1.4\Lambda)^n\beta_0$.

  As stated above, the $n=2$ case is forbidden and a $1/Q^4$ correction is
disfavoured by OPE, but we claim that this argument does not apply to our
$\alpha_S$ at $n=4$, since what we are considering is a possibly universal
correction to $\alpha_S$ arising from OPE, which therefore arises from the
region of small momentum flow.

 \EPSFIGURE[ht]{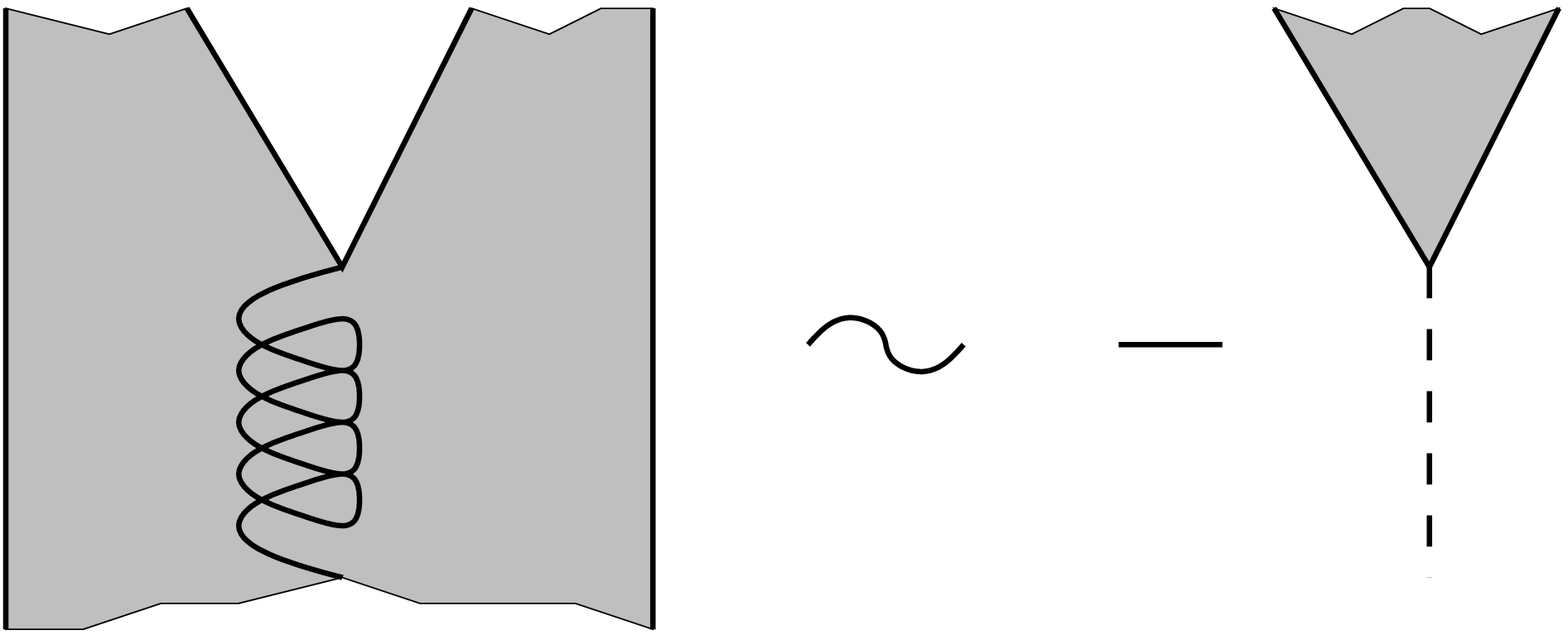, width=12cm}
 {An illustration of the splitting of a `kinked string' by the parton
splitting $g\to q\bar q$.
  Classically, the string `pulling apart' the quark and antiquark
approximates to a string with negative tension `pushing apart' the
partons.
 \label{fig_string}}

  If the $g\to q\bar q$ splitting is indeed enhanced by confinement, an
intuitively appealing picture would be that the quark and the antiquark
are `pulled apart' by the large-distance string-like confining potential,
as shown in fig.~\ref{fig_string}. This would classically approximate to a
string with negative tension `pushing apart' the quark and the antiquark,
and hence by dimensional analysis the contribution to the effective
strength for gluon splitting is expected to be:
 \begin{equation}
  T_\mathrm{eff}=n_fT_R\left[1+
  \mathcal{O}\left(\alpha_S\right)+
  \mathcal{O}\left(\frac{\sigma^2}{Q^4}\right)\right].
 \end{equation}
  $n_fT_R=1$ or 1.5 depending on whether the splitting $g\to s\bar s$ is
allowed. Hence $A\sim\sigma^2$.
  Let us test the validity of this parameter choice.
  First, the peak position in this $\alpha_S$ is given by the point at
which the beta function vanishes in eqn.~(\ref{eqn_nonperturbative_beta}),
i.e.:
 \begin{equation}
  A=\beta_0 Q_\mathrm{peak}^n.
  \label{eqn_aqpeak}
 \end{equation}
  Now from eqn.~(\ref{eqn_qmean}) and $Q_\mathrm{min}=0.4$ GeV,
$Q_\mathrm{peak}=0.575$ GeV. Thus $A=\beta_0Q^2_\mathrm{peak}=0.237$
GeV$^2$ is in good agreement with what one might expect for the string
tension $\sigma\sim0.2$ GeV$^2$ \cite{lattice}. The corresponding 
$\alpha_S$ is shown in fig.~\ref{fig_alphas2}.

  The area underneath $\alpha_S$ and hence $I_0$ in eqn.~(\ref{eqn_sudy})
can be calculated numerically. For $A=(2.3\Lambda)^n\beta_0$ where
$\Lambda=0.25$ GeV such that $Q_\mathrm{peak}=0.575$ GeV, and integrating
up to 0.75 GeV, we obtain $I_0=0.536,0.441,0.389$ and $0.355$ for
$n=2,4,6$ and 8 respectively. From multiplicity considerations the
preferred value is $I_0\sim0.5$ according to the estimation from
eqn.~(\ref{eqn_splitmulti}), such that the values are reasonable, though
the $n=6$ and $8$ cases are possibly too small.

 \EPSFIGURE[ht]{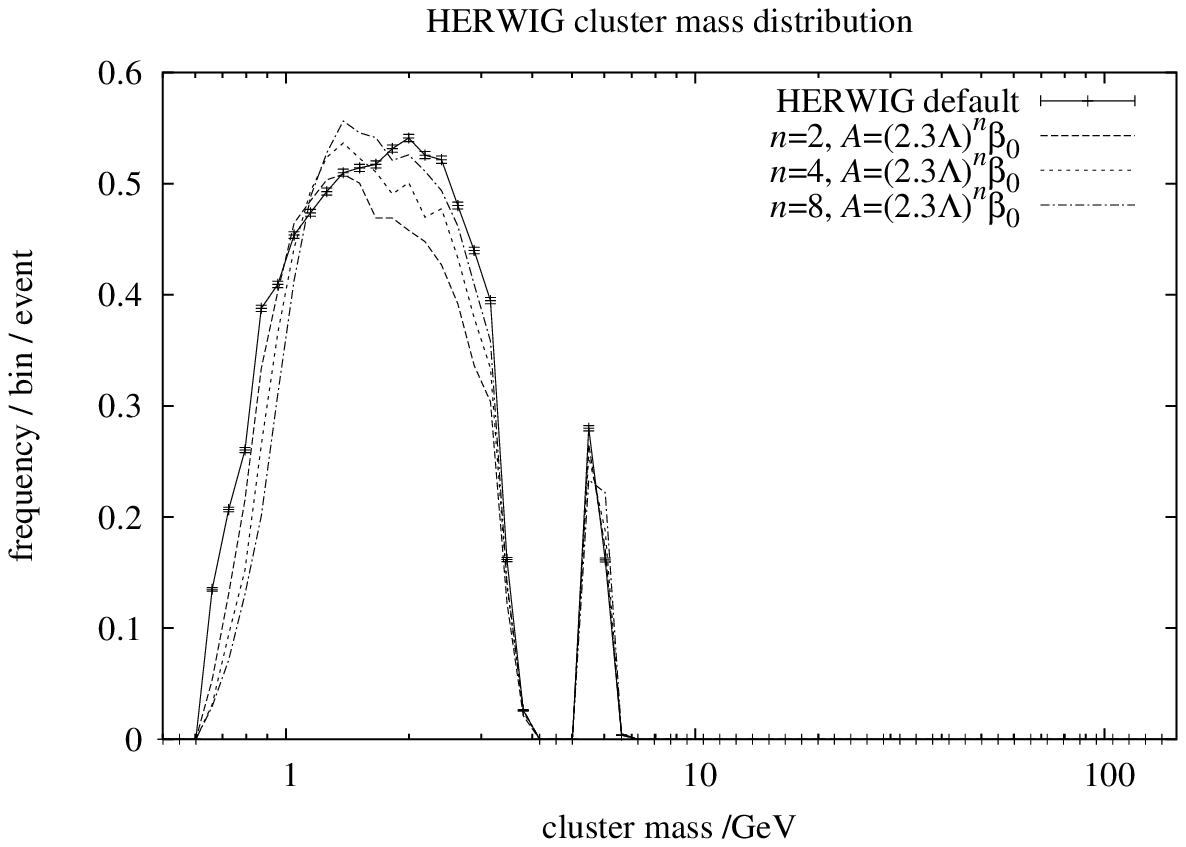, width=14cm}
 {The cluster mass distribution generated using the $\alpha_S$ derived
from power corrected beta function.
 \label{fig_clusmasn8}}

  The cluster mass distribution corresponding to $n=2,4$ and 8 are shown
in fig.~\ref{fig_clusmasn8}. At first sight, it seems that the $n=8$
distribution matches the HERWIG default numbers best, hence implying that
the low-energy tail of the $\alpha_S$ for $n=4$, or 2, is too high. On the
other hand, if the tail is not sufficiently high, as in the $n=8$ case,
there would not be sufficient contribution to $I_0$. Hence there seems to
be a contradiction here.
  However, one reason why the Gaussian approximation in
eqn.~(\ref{eqn_gaussian}) can rectify the small-mass oriented skew of the
cluster mass distribution is that there is some contribution from $Q^2$
above (0.75 GeV)$^2$.
  In HERWIG, the emission cut-off is implemented as being controlled by
the gluon effective mass rather than as having a sharp cut-off, so that
there is inevitably such a contribution.
  The proper study of this, taking account of the mass effects, is beyond
our intention, and is presumably best left to a study incorporating
experimental data, if possible using a generator that is based from the
outset on the modified-$\alpha_S$ prescription rather than a modification
of HERWIG as presented here.

  One way to study the $g\to q\bar q$ dynamics might be to measure the
compensation of the SU(2) light flavour charge.
  This should occur close to the electric charge compensation.

  Let us now turn to a discussion of the analyticity of low energy
$\alpha_S$.

  As stated above, the $\alpha_S$ defined by
eqn.~(\ref{eqn_nonperturbative_alphas}) has complex poles.
  This unwelcome in the sense that it forbids a spectral representation,
viz:
 \begin{equation}
  \alpha_S(Q^2)=-\int_0^\infty\frac{d\mu^2}{\mu^2+Q^2}\rho_S(\mu^2),
  \label{eqn_spectral}
 \end{equation}
  where $\rho_S(\mu^2)$ is the spectral density function. As $\alpha_S$ is
related to the gluon two-point function, the physical interpretation would
be that in the confined vacuum the spectral density associated with the
colour-octet state is ill-defined.
  On the other hand, colour-singlet two-point functions should have a
valid spectral representation such that the complex poles must somehow
cancel.
  We add that in our convention, a physical spectral density function is
negative definite. However, if we impose this condition, we see that
$\alpha_S$ must continue to grow towards low $Q^2$.

  We may rephrase this statement as follows. If colour is confined in the
large distance limit, an $\alpha_S$ that describes the interaction of the
unconfined gluon must vanish in the soft limit. But this is in violation
of the spectral representation if the spectral density function is
negative definite, such that there must be complex poles. These complex
poles must cancel in quantities that are associated with colour singlet
states.

  Let us denote the above $\alpha_S$,
eqn.~(\ref{eqn_nonperturbative_alphas}), that describes gluons both in the
confined and the asymptotically-free vacuum, by $\alpha_S^\mathrm{glu.}$.
  We then write the nonperturbative physics that cancels the poles by
$\alpha_S^\mathrm{had.}$. Then for the overall process of hadronization,
we have:
 \begin{equation}
  \alpha_S^\mathrm{h}=
  \alpha_S^\mathrm{glu.}+\alpha_S^\mathrm{had.}.
  \label{eqn_gluhad}
 \end{equation}
  The former term is universal, but the latter term is not necessarily so.
We also note that from the condition of finite emission probability, we
required the first term to vanish as a power towards low $Q^2$. This is
natural considering that an unconfined gluon can not propagate in the
confined vacuum.
  On the other hand, as the concept of emission probability does not hold
much meaning in the hadronic phase, there is no such requirement governing
the latter term except that it is confined to the region of small $Q^2$,
and it would be reasonable for it to remain non-zero.

  Now let us shift contributions between the two terms such that there is
no complex pole in either of the $\alpha_S$, but the two terms, or more
strictly their moments, are affected minimally, in some sense which we
shall discuss later, for positive and real $Q^2$. Then we can write:
 \begin{equation}
  \alpha_S^\mathrm{h}=
  \alpha_S^\mathrm{conf.}+\alpha_S^\mathrm{decay}.
  \label{eqn_confdec}
 \end{equation}
  The former `confinement' term is still at least almost universal.
  $\alpha_S^\mathrm{decay}$ is expected to be the contribution to the
observable under attention from the decay of the nonperturbative clusters
and hadrons, and is expected to be effective at a lower scale than the
former term.

 \TABULAR[p]{|c|c|c|c|}{
  \hline scheme       & $F_{1,1}$ & $F_{2,1}$ & $F_{3,1}$ \\\hline
  $n=2$               & 0.378     & 0.394     & 0.381     \\
  $n=4$               & 0.394     & 0.419     & 0.401     \\
  $n=6$               & 0.390     & 0.421     & 0.404     \\
  $n=8$               & 0.385     & 0.420     & 0.404     \\\hline
  Webber's $\alpha_S$ & 0.511     & 0.450     & 0.410     \\\hline}
 {First three moments, $F_{p,1}$(2 GeV), of low energy $\alpha_S$.
 \label{tab_moments}}

  Using $\alpha_S^\mathrm{h}$ which has a dispersive representation, as do
its two constituents $\alpha_S^\mathrm{conf.}$ and
$\alpha_S^\mathrm{decay}$, we can calculate power corrections following
the procedure of ref.~\cite{dispersive}. When doing so, the quantities of
main interest are the first few moments of $\alpha_S$, i.e., $F_{p,1}(Q)$,
where $F_{p,q}(Q)$ \cite{webber_alphas} are defined by:
 \begin{equation}
  F_{p,q}(Q)=\frac{p}{Q^p}
  \int_0^Q\frac{dk}{k}k^p
  \left[\alpha_S(k^2)\right]^q.
  \label{eqn_fpqq}
 \end{equation}

  $\alpha_S^\mathrm{glu.}$ has complex poles and therefore can not be used
to calculate power corrections directly in the dispersive approach.
  However, if our argument above is correct, we can choose
$\alpha_S^\mathrm{conf.}\sim\alpha_S^\mathrm{glu.}$.
  As these moments can be calculated numerically for any choice of
$\alpha_S$ so long as it is integrable, it then follows that one may
calculate them for $\alpha_S^\mathrm{glu.}$ as an estimate of the moments
of $\alpha_S^\mathrm{conf.}$.
  We take the $n=2,4,6$ and 8 cases with $A$ set to
$(2.3\Lambda)^n\beta_0$ and compare the result with $\alpha_S$ of
ref.~\cite{webber_alphas}. For $Q=2$, the result is shown in
tab.~\ref{tab_moments}.

  The $\alpha_S$ of ref.~\cite{webber_alphas} is defined to describe event
shapes and is therefore an $\alpha_S^\mathrm{h}$.
  Physically we would expect that both $\alpha_S^\mathrm{conf.}$ and
$\alpha_S^\mathrm{decay}$ would be positive definite. If so, we have
$\alpha_S^\mathrm{h}>\alpha_S^\mathrm{conf.}$. A more important criterion
is that this is satisfied in the moments. We see from
tab.~\ref{tab_moments} that this is indeed the case.
  Furthermore, we see that $\alpha_S^\mathrm{decay}$ contribution only 
accounts for $\mathcal O(20\%)$ of the contribution to the moments.

  From the requirement that $\alpha_S^\mathrm{decay}$ mainly affects the
region of $Q^2$ lower than $\alpha_S^\mathrm{conf.}$, or more strictly
$\alpha_S^\mathrm{glu.}$, we expect that the higher moments of
$\alpha_S^\mathrm{h}$ and $\alpha_S^\mathrm{conf.}$ are similar. This is
also satisfied in tab.~\ref{tab_moments}. However, the similarity is
exaggerated by the choice of the cut-off scale, 2 GeV, which is high
compared with the hadronic scale.

  This argument governing the moments also implies that the moments
$F_{p,1}(Q)$ derived from experimental data must be greater than the
contribution from $\alpha_S^\mathrm{conf.}$, such as those listed in
tab.~\ref{tab_moments} except the one in the bottom row. From the data
compiled in ref.~\cite{universality}, which we reproduce in
tab.~\ref{tab_gavin}, $F_{1,1}(2$ GeV) is in the range 0.391 to 0.560.
Although 0.391 measured in the jet broadening observable $B_W$ is quite
low, it is not in contradition of this statement.

 \TABULAR[p]{|c|c|c|c|}{
  \hline Variable & $F_{1,1}$(2 GeV)\\\hline
  $B_T$      &$0.4508\pm0.0225$\\\hline
  $B_W$      &$0.3911\pm0.0305$\\\hline
  $1-T$      &$0.4976\pm0.0087$\\\hline
  $C  $      &$0.4527\pm0.0110$\\\hline
  $M^2_h/Q^2$&$0.5602\pm0.0224$\\\hline}
 {Moment $F_{1,1}$(2 GeV) of low energy $\alpha_S$, obtained by
 fit to event shape variables. Numbers taken from
ref.~\cite{universality}.
 \label{tab_gavin}}

  Let us proceed to estimating the ambiguity of $F_{1,1}$ in the process
of analytization in between eqns.~(\ref{eqn_gluhad}) and
(\ref{eqn_confdec}).
  We can do this by simply subtracting off the complex poles in
$\alpha_S^\mathrm{glu}$ and calculating the corresponding contribution to 
$F_{1,1}$.
  This analytization procedure is equivalent to that adopted in
ref.~\cite{analytization}.

  We first define for $u$ and $v$ real:
 \begin{equation}
  u+iv = \frac{A}{\beta_0Q^n}.
 \end{equation}
  Then $Q^2/\Lambda^2$ is given by one of the roots:
 \begin{equation}
  \frac{Q^2}{\Lambda^2}=
  \left(\frac{A/(\beta_0\Lambda^n)}{u+iv}\right)^{2/n}
  \equiv \left(\frac{a}{u+iv}\right)^{1/\nu},
  \label{eqn_qoverlam}
 \end{equation}
  where we defined for convenience $a=A/(\beta_0\Lambda^n)$ and $\nu=n/2$.
  Let us specialize in the case of even $n$. It turns out that out of the
$\nu$ roots of eqn.~(\ref{eqn_qoverlam}), the only relevant one is the one
whose phase is $1/\nu$ times the phase of $u-iv$, where this phase is 
taken to be between $-\pi$ and $+\pi$.

  At the pole we have:
 \begin{equation}
  \alpha_S^{-1} =
  \beta_0\log\left(\frac{Q^2}{\Lambda^2}\right)+\frac{2A}{nQ^n}=
  \beta_0\left[\log\left(\frac{a}{u+iv}\right)^{1/\nu}+
  \frac{u+iv}{\nu}\right]=0.
  \label{eqn_aseqzero}
 \end{equation}
  Hence:
 \begin{equation}
  u+iv = a\exp\left(u+iv\right),
  \label{eqn_upluiv}
 \end{equation}
  from which we derive:
 \begin{equation}
  u=\frac{v}{\tan v},\quad a=\frac{v}{e^u\sin v}.
  \label{eqn_uanda}
 \end{equation}
  As $a$ is positive, we require that $v/\sin v$ is also positive. With
this condition, the contour of poles in the $u+iv$ space is plotted in
fig.~\ref{fig_contour}. $a$ tends to $+0$ as $u\to+\infty$ and $+\infty$
as $u\to-\infty$.

 \EPSFIGURE[ht]{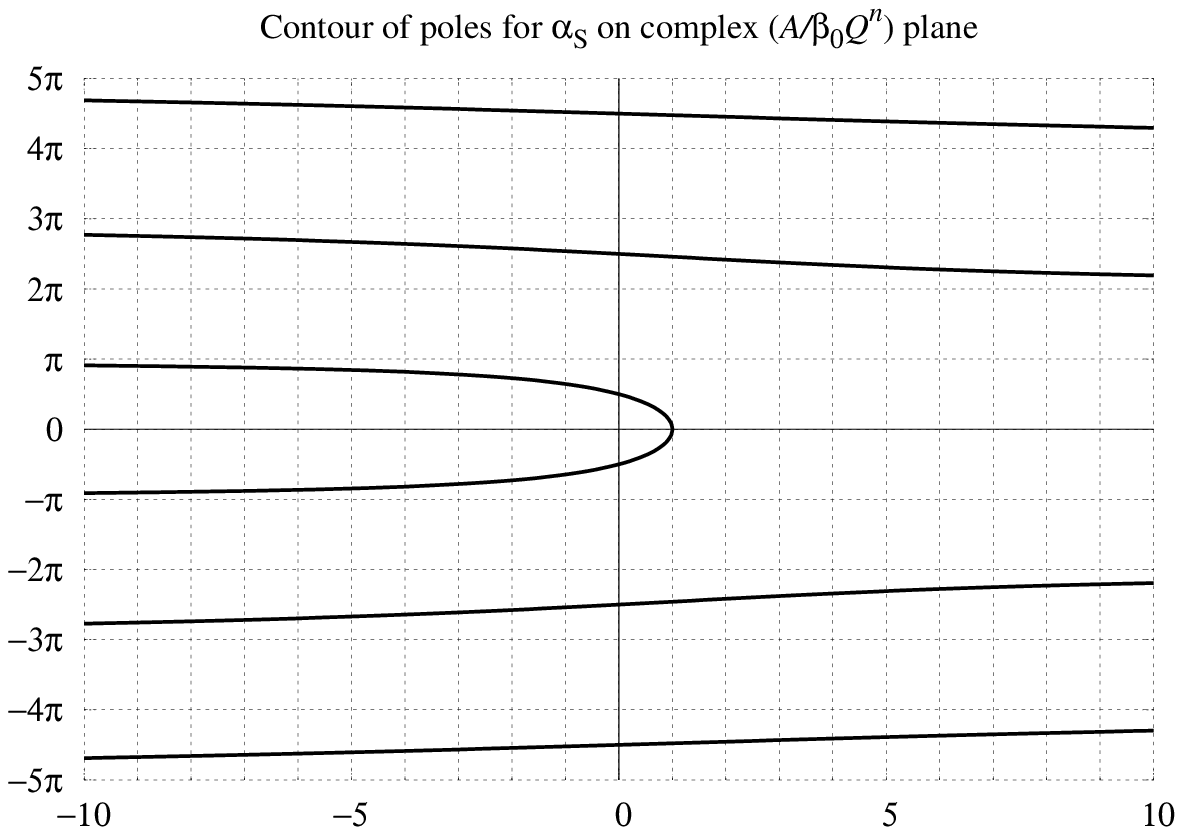, width=14cm}
 {The contour of poles for $\alpha_S^\mathrm{glu.}$ on the complex 
$A/(\beta_0Q^n)$ plane.
 \label{fig_contour}}

  Eqns.~(\ref{eqn_uanda}) are consistent with eqn.~(\ref{eqn_aseqzero})  
only if:
 \begin{equation}
  |v|<\nu\pi.
 \end{equation}
  Thus for $n=2,4$, there are two solutions which are complex conjugates,
whereas for $n=6,8$ there are four solutions, and so on. We note
furthermore that the lowest solution is complex only if $a>e^{-1}$.

  The poles of $\alpha_S$ can be found for given $A$ and hence $a$ by
solving eqns.~(\ref{eqn_uanda}) to obtain $v$ and $u$ in terms of $a$. The
residue on the $Q^2$ plane is given in terms of the beta function
evaluated at the pole, $\beta=\beta_0(1-u-iv)$, as:
 \begin{equation}
  \mathrm{residue} =
  \frac{Q^2}{\beta\left(Q^2\right)}\Biggr|_{Q^2\mathrm{\ at\ pole}}.
  \label{eqn_residue}
 \end{equation}

  When there is no complex pole, the spectral representation of
eqn.~(\ref{eqn_spectral}) has the well-known solution in terms of the 
discontinuity of $\alpha_S$ on the negative real axis:
 \begin{equation}
  \rho_S(\mu^2)=\frac1{2\pi i}
  \left[\alpha_S(\mu^2e^{i\pi})-\alpha_S(\mu^2e^{-i\pi})\right].
  \label{eqn_discontinuity}
 \end{equation}
  As stated earlier, a physical $\rho_S$ is negative definite in our
convention.
  The generalization of eqns.~(\ref{eqn_spectral}) and
(\ref{eqn_discontinuity}) to include contribution from the complex poles
yields:
 \begin{equation}
  \alpha_S(Q^2)=-\int_0^\infty\frac{d\mu^2}{\mu^2+Q^2}\rho_S(\mu^2)-
  \sum_\mathrm{pole}
  \left[\frac1{\beta(\mu^2)\left(1-Q^2/\mu^2\right)}\right]
  \Biggr|_{\mu^2\mathrm{\ at\ pole}},
  \label{eqn_spectral_modified}
 \end{equation}
  This general expression does not depend on the particular form of
$\alpha_S$ but assumes that the complex singularities are all simple
poles. $\alpha_S$ can be analytized by subtracting the complex poles, or
by adding:
 \begin{equation}
  \delta\alpha_S(Q^2)=\sum_\mathrm{pole}
  \left[\frac1{\beta(\mu^2)\left(1-Q^2/\mu^2\right)}\right]
  \Biggr|_{\mu^2\mathrm{\ at\ pole}}.
  \label{eqn_deltaas}
 \end{equation}
  Applying the formulae to the $\alpha_S$ in
eqn.~(\ref{eqn_nonperturbative_alphas}), the resultant analytized
$\alpha_S$ has a negative definite spectral density, given by:
 \begin{equation}
  \rho_S(\mu^2)=-\frac1{\beta_0}
  \left[\left[\log\left(\frac{\mu^2}{\Lambda^2}\right)
  +\frac{2A}{n\beta_0\mu^n}\right]^2+\pi^2\right]^{-1}.
  \label{eqn_discontinuity_nonperturbative_alphas}
 \end{equation}
  $n$ is assumed to be even. Denoting the analytized $\alpha_S$ by
$\overline{\alpha_S}$, as $\overline{\alpha_S}$ continues to grow with
decreasing $Q^2$, $\overline{\alpha_S}(0)$ is a finite number, which can
be calculated either by integrating
eqn.~(\ref{eqn_discontinuity_nonperturbative_alphas}), or more simply by
summing eqn.~(\ref{eqn_deltaas}) at $Q^2=0$, i.e., summing $1/\beta$ over
the poles.

  $\overline{\alpha_S}(Q^2)$ has an unphysical $1/Q^2$ behaviour at large
$Q^2$, which must be cancelled by additional contribution from $\rho_S$.
This additional contribution can be defined as a negative definite
quantity only if the sum of the residues $Q^2/\beta$,
eqn.~(\ref{eqn_residue}), is positive. By evaluating
eqn.~(\ref{eqn_residue}) in the $A\to\infty$ limit, we see that this is
always satisfied when $\nu=n/2$ is even, but violated for odd $\nu$ when
$A$ is large.

  The ambiguity of $F_{1,1}$ is estimated by integrating
eqn.~(\ref{eqn_deltaas}) according to eqn.~(\ref{eqn_fpqq}). We obtain
$\delta F_{1,1}=0.071,0.016$ and $-0.052$ respectively for $n=2,4$ and 8.
$A$, as before, is given by $(2.3\Lambda)^n\beta_0$.

  We see that the ambiguity, amounting to up to $\sim10\%$ of $F_{1,1}$,
is under control. Had we started from the unmodified perturbative
$\alpha_S$ \cite{analytization}, the corresponding $\delta F_{1,1}$ would
be undefinable.

  As the procedure followed here gives an estimate only, it is not
possible to draw definitive conclusions based on these numbers, but
comparison with tabs.~\ref{tab_moments} and \ref{tab_gavin} shows that the
magnitude of $\delta F_{1,1}$ is consistent with data, and in our opinion
$F_{1,1}+\delta F_{1,1}$, corresponding to $\overline{\alpha_S}$, also
does not show disparity with tab.~\ref{tab_gavin} that renders any 
particular choice of $n$ completely unrealistic.

  We have so far calculated the confinement contribution of
eqn.~(\ref{eqn_confdec}) to $F_{1,1}$ and the ambiguity due to 
analytization in between eqns.~(\ref{eqn_gluhad}) and (\ref{eqn_confdec}).
  It is not possible to calculate the decay contribution within our 
present framework.
  A very naive expectation would be that since the decay process occurs at
lower energies than the confinement process, the highest energy component
of $\alpha_S^\mathrm{had.}$ would be given by the same $\delta\alpha_S$
that cancels the complex pole in $\alpha_S^\mathrm{glu.}$. If we accept
this, we would expect that although $\alpha_S^\mathrm{had.}$ is
non-universal, its size is estimated by $\delta F_{1,1}$ that we have
calculated above. Again, all cases are consistent with data and is in
agreement with the small size of the $\alpha_S^\mathrm{had.}$ contribution
to $F_{1,1}$ seen in tab.~\ref{tab_moments}, or in the difference between
tabs.~\ref{tab_moments} and \ref{tab_gavin}.

  One may ask how our results depend on the input parameters. We recall
that we determined our value of $A$ by tuning the peak position of
$\alpha_S$, as in eqn.~(\ref{eqn_aqpeak}). We may shift the peak position
slightly and see how this affects our prediction.
  For $n=4$, when $Q_\mathrm{peak}$ is modified from $0.575$ GeV
corresponding to $Q_\mathrm{min}=0.4$ GeV to $0.5$ GeV corresponding to
$Q_\mathrm{min}=0.25$ GeV, which we consider to be unreasonably small,
$\delta F_{1,1}$ decreases from $0.015$ to $-0.004$, while $F_{1,1}$ from
integrating $\alpha_S^\mathrm{glu.}$ itself increases from $0.394$ to
$0.424$.
  We have also calculated the case $Q_\mathrm{peak}=0.65$ GeV such that
$Q_\mathrm{min}=0.55$ GeV which again is unreasonable. This yields $\delta
F_{1,1}=0.032$ whereas $F_{1,1}$ is reduced to $0.369$. We observe that
$F_{1,1}+\delta F_{1,1}$ is less sensitive to $A$ than $F_{1,1}$ itself
is, and in any case the dependence is small enough to trust the numbers
calculated using the procedures presented here.

 \section{Conclusions}

  We have reconsidered the phenomenon of local charge compensation from
the viewpoint of colour-preconfined models of hadronization.

  We have suggested that the study of minijet charge as a function of the
resolution parameter $y_\mathrm{cut}\propto {k_T^2}_\mathrm{cut}$ could
provide information on the nature of dynamics governing hadronization. In
particular, the scale at which local charge compensation is maximally
violated, i.e., where the minijet charge is peaked, is the scale at which
nonperturbative, confinement, dynamics sets in.
  We have demonstrated by simulations using HERWIG that modifications in
the parameters governing hadronization could lead to significant
differences in the behaviour of the minijet charge.
  We have made a comparison with PYTHIA and found that the predictions of
the two generators are similar.

  The values for minijet charge depend on the scheme used for clustering
minijets. Our default procedure uses the Cambridge algorithm, but we have
demonstrated that the combination of the JADE algorithm in the confined
region and angular-ordered Durham algorithm in the perturbative region
results in less contamination from misidentified tracks. We have proposed
a `preclustered' scheme which shows marked improvement compared with the
other algorithms when the preclustering scale $y_\mathrm{had}$ is tuned to
(0.5 GeV)$^2$.

  In addition to the $e^+e^-$ case, we have presented a simple analysis of
the minijet charge at hadron collisions and found that due to the extra
contamination in this case, the peak in the minijet charge is weakened.
The peak disappears completely when the soft underlying events are added 
according to the default option of HERWIG.

  We have studied a modification of the HERWIG default procedure to
express the cluster-splitting dynamics at the end of the parton shower
phase in terms of emissions that are considered unresolved in the course
of the ordinary parton shower. Through this analysis, we have found that
the lowest scale of cluster-splitting dynamics is about 0.4 GeV.

  This analysis also suggests that because the product of neighbouring
cluster masses is smaller in this case compared with the HERWIG default
procedure, the peak height in the minijet charge is necessarily smaller
than is suggested by HERWIG.
  On the other hand, the relaxation of HERWIG's sudden transition to the
confinement phase is expected to smear this peak.

  Our simulations for this part of the study were carried out using a
simple modification of HERWIG with several approximations.
  It is hoped that simulation using Monte-Carlo event generators of the
next generation, that are based from the outset on the modified-$\alpha_S$
prescription, will also be available in the future.
  We emphasize that comparison with experiment can nevertheless be made
using our present approach to yield useful information that has less
theoretical prejudice,
  but once a particular model of $\alpha_S$ is available, such as the one
proposed in this paper, simulation based from the outset on
modified-$\alpha_S$ could leave less ambiguity in the theoretical
interpretation.

  We have discussed the possible implications of our findings to the
underlying nonperturbative dynamics.
  In particular, we studied a model in which the transition between the
perturbative phase and the nonperturbative phase during the course of
hadronization is driven by the semi-perturbative splitting $g\to q\bar q$,
possibly enhanced by the string-like confining force.
  This gives rise to an $\alpha_S$ which vanishes as a power towards low
$Q^2$ but has complex poles.
  We have argued, in agreement with available data, that the moments of
this $\alpha_S$ gives an estimate of the part of the power corrections to
event shapes that is universal and comes from soft gluon emission, whereas
its complex poles give an estimate of the ambiguity, which is found to be
small.
  In this picture, the remaining part of power corrections is related to
the cascade decay of the nonperturbative, confined, clusters, and in our
present framework can only be estimated to be of the same order as the
ambiguity of the former part.

 \acknowledgments

  The author thanks Bryan Webber for inspiring and extensive discussions.

 \end{document}